\documentclass[
amsmath,amssymb,
aps,physrev,pra,
twocolumn
]{revtex4-2}

\usepackage[version=3]{mhchem}
\usepackage{graphicx}
\usepackage{dcolumn}
\usepackage{bm}
\usepackage{multirow}
\usepackage{tabularx}
\usepackage{booktabs}
\usepackage{xr-hyper}

\begin{document}
\title{\textbf{Accurate representation of the continuum orbital in a Gaussian basis set: application to Auger decay rates}}

\author{Dibyendu Mahato}
\author{Saikat Roy}
\author{Wojciech Skomorowski}
\email{w.skomorowski@cent.uw.edu.pl}
\affiliation{Centre of New Technologies, University of Warsaw, 02-097 Warsaw, Poland.}
\begin{abstract}
Many processes in chemistry and physics, such as autoionization, photoionization, or electron-impact collisions, involve unbound electrons.
Despite their importance, theoretical methods for modeling continuum electrons remain significantly less developed than those for purely bound electronic states. 
Here we present a novel approach for treating continuum electrons within Gaussian-based quantum chemistry. 
The method represents continuum orbital in a discrete basis of standard Gaussian-type orbitals (GTOs), where the continuum wave function is obtained by solving the Lippmann--Schwinger equation with proper boundary conditions using an analytical representation of the Green’s function in an auxiliary GTO basis set. The resulting continuum orbital is multicentered and multichannel, providing a rigorous description of scattering states. As an initial application of the new methodology, we compute Auger decay rates for several atomic and molecular systems, including Ne, H$_2$O, CO$_2$, and Ar .
We demonstrate a consistent improvement of the results with increasing the quality of the continuum representation. 
We show also that higher-order electron-correlation effects in the bound domain are essential for an accurate treatment of satellite transitions. The present formalism therefore provides a systematically improvable Gaussian-based framework for continuum-electron problems that is fully compatible with standard quantum chemistry methods. 
\end{abstract}

\maketitle

\section{Introduction}
Quantum chemistry has witnessed tremendous progress over the past decades, enabling \textit{ab initio} calculations for systems containing even millions of electrons~\cite{wieners2026quantum}. A major factor behind this progress has been the widespread use of Gaussian basis functions to represent electronic many-body states, offering an excellent compromise between computational efficiency in evaluating multicenter electron integrals and accuracy in describing electronic densities. However, these developments have focused almost exclusively on bound-state electronic structure problems, where wave functions are square-integrable and spatially localized. Considerably less attention has been devoted to the \textit{ab initio} description of unbound electronic states, which require fundamentally different asymptotic boundary conditions. Consequently, continuum-electron methods remain much less developed within standard quantum chemistry, despite their importance in numerous chemical and physical processes. Their relevance has grown further with advances in experimental techniques, particularly those employing high-energy light sources such as X-ray free-electron lasers\cite{ullrich2012free, pellegrini2016physics, Gelmukhanov2021resonant, rudenko2017femtosecond, Berrah2025tut}. These developments create a growing need for efficient and accurate continuum-electron methods compatible with modern quantum chemistry.

From this perspective,  Gaussian basis functions are particularly attractive for describing continuum electrons because they enable seamless integration of continuum phenomena with correlated electronic-structure methods developed for bound-state problems.
Existing Gaussian-based approaches to continuum-electron problems can be broadly divided into two classes. In the first, the continuum orbital is represented explicitly as an expansion in a suitably designed Gaussian basis\cite{Kaufmann1989cont, nestmann1990optimized, faure2002gtobas,Coccia2016cont,Nana2025cont}. Several methods of this type have been proposed. 
Among them, the Schwinger variational method\cite{lucchese1986applications} is particularly well suited to Gaussian basis sets because its formulation can be expressed entirely in terms of square-integrable functions, without requiring the basis itself to reproduce the long-range asymptotic behavior of the scattering wavefunction. As a result, Gaussian-based implementations of the Schwinger variational method have become well-established for treating photoionization and electron–molecule scattering problems\cite{Lucchese1980,Lucchese1980ion,Lucchese1981,Fliflet1978, stratmann1996multiplet}. Other scattering approaches, including the $R$-matrix\cite{schneider1975r, tennyson2010electron, carr2012ukrmol} and complex Kohn methods\cite{schneider1988complex, miller1987new}, have also been implemented using Gaussian basis sets. However, unlike the Schwinger variational approach, these methods cannot provide the scattering solution using only square-integrable trial functions. In the $R$-matrix method, the Gaussian expansion is restricted to the inner region, while the outer-region solution is obtained separately. Similarly, the complex Kohn method supplements the Gaussian expansion with explicit continuum channel functions that enforce the correct asymptotic boundary conditions. In the context of Auger decay, Colle and Simonucci\cite{Colle1988,Colle1989,Colle1990} proposed a Gaussian-based approach in which the continuum orbital is obtained by directly solving the Lippmann–Schwinger equation.


The second class of Gaussian-based continuum methods avoids an explicit representation of the continuum orbital, remaining entirely within \(L^2\)-integrable basis functions. In these approaches, continuum properties are obtained indirectly using additional techniques, such as Stieltjes imaging, which enables to extract physical obervables from discretized approximation to the continuous spectrum. Such methods have proven particularly successful for photoionization\cite{Carravetta1987,cacelli1988h2s,gokhberg2009molecular}, Auger decay\cite{Carravetta1987,Kolorenc2011}, and interatomic Coulombic decay\cite{averbukh2005ab,Kolorenc2020}. Similarly, non-Hermitian methods based on complex scaling or complex basis functions operate exclusively within \(L^2\)-integrable Gaussian basis-set representation and have been widely applied to model metastable electronic states and Auger spectra in small- to medium-sized molecules\cite{mccurdy1978extension,moiseyev1998quantum,Zuev2014,Matz2022}.

Alternatively, continuum orbital can be represented numerically rather than through Gaussian basis expansion.
In such approaches, the continuum wave function is typically obtained directly from the numerical solution of the coupled differential Schr\"odinger equations as in single-center expansion method\cite{demekhin2011single, novikovskiy2019singleceter} or with spherically averaged molecular potential\cite{grell2019multireference, grell2020multi}. Intermediate between purely Gaussian and fully numerical representations are approaches based on B-spline basis functions (or hybrid Gaussian-B-spline basis), which have found broad applications in continuum-electron calculations\cite{brosolo1992variational, toffoli2002convergence, marante2014hybrid, marante2017hybrid, Zapata2024bsp}. In this case, the continuum problem is solved within a finite spatial region large enough to reach the asymptotic regime. The properly normalized scattering wave function, or equivalently the \(K\)-matrix, can then be extracted by matching the asymptotic part of the numerical solution to appropriate combinations of regular and irregular asymptotic functions. However, despite their proven accuracy for describing continuum states, numerical and B-spline continuum representations are generally less straightforward to combine with standard Gaussian-based electronic-structure methods, which benefit heavily from analytical Gaussian integrals algorithms.

In the present work, we develop a continuum-electron methodology that combines key advantages of both explicit scattering formulations and pure Gaussian-based electronic-structure theory. The central idea is to construct continuum orbital through the Lippmann--Schwinger equation \cite{Lippmann1950} with the correct scattering boundary conditions imposed explicitly, while representing all quantities in a finite Gaussian basis. In practice, this requires efficient evaluation of matrix elements involving the free-particle (or Coulomb) Green’s function in multicenter Gaussian basis sets. To facilitate this step, we recently derived analytical expressions for Green’s-function matrix elements in bases of real spherical Gaussian functions and plane-wave--modulated Gaussian functions \cite{Mahato2026}. 
The proposed methodology naturally provides properly normalized continuum orbital together with direct access to channel-resolved observables and angular distributions of unbound electron. At the same time, the Gaussian representation allows seamless integration with correlated bound-state methods and standard finite-basis electronic-structure implementations. An important aspect of the present framework is the explicit inclusion of interchannel coupling effects, which enables simultaneous treatment of electron correlation both within the bound-state domain and in the continuum coupling between different decay channels. The present formalism has been implemented within the \textsc{Q-Chem}\cite{QChem5} quantum chemistry package and employs equation-of-motion coupled-cluster theory restricted to single and double excitations (EOM-CCSD) for the description of electron correlation in the bound domain\cite{Stanton1993, Bartlett2012}. However, the methodology itself is formulated in a general manner and can be combined with a broad range of correlated electronic-structure methods.

The present approach is particularly well suited for localized continuum-related phenomena, such as Auger decay, where the essential contribution from the continuum orbital is restricted to the short-range interaction region occupied by the bound electrons. Therefore, as an initial application, we apply the formalism to calculations of Auger decay rates in several benchmark atomic and small molecular systems, including K-shell Auger decay in \ce{Ne}, \ce{H2O}, and \ce{CO2}, as well as L-shell Auger decay in \ce{Ar}.

The calculation of Auger decay rates in atoms and molecules has a long history, and a variety of theoretical approaches has been developed for treating continuum electrons in such processes. As discussed above, these methods can be broadly classified according to whether the continuum orbital is represented explicitly\cite{faegri1979calculated, Colle1990, grell2019multireference, Inhester2012, zahringer1992molecular} or implicitly\cite{Nicolaides1985, Kolorenc2011, carravetta2000ab, Liegener1983, schimmelpfennig1995ab}.
In addition, several approximations have been proposed to reduce the computational cost of continuum-electron calculations, with the one-center approximation (OCA)\cite{Siegbahn1975, Fink1995, tenorio2022multireference} perhaps being the most successful one. The present work builds directly upon earlier studies, in which simplified continuum models based on plane wave and Coulomb wave were employed to approximate state of the emitted Auger electron\cite{Skomorowski2021, Skomorowski2021_2}. While such approximations can provide qualitatively reasonable results, they generally lack systematic accuracy and predictive reliability. Here we demonstrate that a more accurate description of the continuum electron, combined with explicit treatment of electron correlation both within the bound-state sector and through interchannel coupling, leads to systematic improvement in the calculated Auger decay widths. To make the comparison fully meaningful and controlled, we focus primarily on benchmarking against approaches that employ the same underlying bound-state electronic-structure methodology based on EOM-CCSD theory. This allows us to isolate the effects arising specifically from improvements in the continuum representation, higher-order correlation effects in the bound-state sector, and inter-channel coupling. Where available, we also compare our results with calculations employing continuum orbital obtained from numerical solution of the Schr\"odinger equation, which can serve as a high-reference standard for assessing the quality of the continuum description.

The paper is organized as follows. Section~II presents the theoretical methodology for constructing multicenter, multichannel continuum orbital using the Lippmann--Schwinger equation and Gaussian-type basis functions. Section~III presents benchmark Auger decay calculations for K-shell-ionized \ce{Ne}, \ce{H2O}, and \ce{CO2}, and L-shell-ionized \ce{Ar}. We analyze the effects of improved continuum representation, interchannel coupling, and higher-order electron-correlation effects on satellite transitions, and compare the results with relevant experimental and theoretical data. Section~IV summarizes the conclusions and outlines future applications.

\section{Theory}
We consider many-electron states describing the interaction of a single unbound electron with a bound target. Although the formalism presented below is general, a representative example is provided for the Auger decay process, in which a highly-excited core-hole electronic state decays through the emission of a single electron. Such a process can be represented as
\begin{equation}
\label{eq:1}
[\mathrm{A}^{*}]^{+(Z_{\mathrm{net}}-1)} \;\rightarrow\; \mathrm{A}_{n}^{+Z_{\mathrm{net}}} + e^-,
\end{equation}
where $Z_{\mathrm{net}}$ denotes the net charge of the stable ionic product. 
Typical values of $Z_{\mathrm{net}}$ are 2 for normal Auger decay, 1 for resonant Auger decay or photoionization of a neutral target, and 0 for electron scattering from a neutral target. The parameter $Z_{\mathrm{net}}$ is kept general to emphasize that the formalism is not restricted to autoionization and applies more broadly to processes involving continuum electron.

\subsection{Single-channel Lippmann--Schwinger formulation}
We begin with the single-channel approximation, in which the stable $(N-1)$-electron ionic product $\mathrm{A}_n^{+Z_{\mathrm{net}}}$ is restricted to a single electronic state, denoted by $\Psi_n^{N-1}$. The $N$-electron wave function describing  the products on the right-hand side of Eq.~(\ref{eq:1}) is then represented as
\begin{equation}
\label{eq:2}
\Psi_{\mathbf{k}_n}^{(\pm), N}(\mathbf r_1,\ldots,\mathbf r_{N})=
\hat{\mathcal A}\left\{\phi_{\mathbf{k}_n}^{(\pm)}(\mathbf r_1)\,\Psi_{n}^{N-1}(\mathbf r_2,\ldots,\mathbf r_{N})\right\}
\end{equation}
where $\phi_{\mathbf{k}_n}^{(\pm)}$ is the continuum orbital associated with ionic channel $n$, and $\hat{\mathcal A}$ is the antisymmetrization operator. The superscript $(\pm)$ denotes outgoing $(+)$ or incoming $(-)$
scattering boundary conditions. The target states $\Psi_n^{N-1}$ are assumed to be orthonormal eigenfunctions of the $(N-1)$-electron Hamiltonian,
\begin{equation}
\hat H^{N-1} \Psi_n^{N-1} = E_n \Psi_n^{N-1},
\qquad
\langle \Psi_m^{N-1} | \Psi_n^{N-1} \rangle = \delta_{mn}.
\end{equation}
The kinetic energy of the continuum electron in channel $n$ is then determined by energy conservation as
\begin{equation}
\varepsilon_n = \frac{k_n^2}{2} = E-E_n.
\end{equation}
where $k_n=|\mathbf{k}_n|$ is the magnitude of its asymptotic wave vector, $E$ is the total energy of the system, and $E_n$ is the threshold energy of channel $n$. To simplify notation, we omit the spin coordinates and spin-adaptation coefficients in Eq.~(\ref{eq:2}). Since the electronic Hamiltonian considered here is spin independent, both the total wave function $\Psi_{\mathbf{k}_n}^{(\pm), N}$ and the target states $\Psi_n^{N-1}$ are eigenfunctions of $\hat{S}^2$ and $\hat{S}_z$ operators. 
Consequently, the spin-adapted wave function can be constructed straightforwardly by coupling the target and continuum-electron spin functions with the appropriate Clebsch--Gordan coefficients; see, for example, Eq.~(19) of Ref.~\citenum{Skomorowski2021} and Eqs.~(1)--(3) of Ref.~\citenum{Inhester2012} for explicit expressions for the most common total-spin symmetries ($S=0$, 1, and 2).

In the next step we consider the full $N$-electron Schrödinger equation,
\begin{equation}
\hat{H}^{N} \Psi_{\mathbf{k}_n}^{(\pm), N} = E\Psi_{\mathbf{k}_n}^{(\pm), N},
\label{eq:3a}
\end{equation}
where $\hat{H}^{N}$ is the Hamiltonian of the $N$-electron system.
Separating the coordinate $\mathbf r_1$ of the continuum electron, the $N$-electron Hamiltonian can be written as
\begin{equation}
\hat{H}^{N} = \hat{H}^{N-1} -\frac{1}{2}\nabla_1^2 -\sum_A \frac{Z_A}{|\mathbf r_1-\mathbf R_A|} + \sum_{i=2}^{N}\frac{1}{|\mathbf r_1-\mathbf r_i|},
\label{eq:3b}
\end{equation}
where $Z_A$ and $\mathbf R_A$ denote the charge and position of nucleus $A$, respectively.
The full $N$-electron channel states are normalized according to the continuum normalization condition
\begin{equation}
\left\langle \Psi_{\mathbf{k}_n}^{(\pm), N} \middle| \Psi_{\mathbf{k}'_m}^{(\pm), N} \right\rangle = \delta_{nm}\, \delta^{(3)} \left(\mathbf{k}_n-\mathbf{k}'_{n}\right),
\label{eq:continuum_norm}
\end{equation}
where the Kronecker delta reflects the orthogonality of the ionic target states, while the Dirac delta expresses the continuum normalization within a given channel. Substituting the ansatz of Eq.~(\ref{eq:2}) into the full $N$-electron Schrödinger equation, Eq.~(\ref{eq:3a}), and projecting the resulting equation onto the target ionic state $\Psi_n^{N-1}$ yields the following effective equation for the continuum orbital $\phi_{\mathbf{k}_n}^{(\pm)}$ in the single-channel approximation (in atomic units):
\begin{equation}
\label{eq:se_schroedinger2}
\left[-\frac{\nabla^2}{2}+\hat{V}_{n n}(\mathbf r)\right] \phi_{\mathbf{k}_n}^{(\pm)}(\mathbf r) =\frac{k_n^2}{2} \, \phi_{\mathbf{k}_n}^{(\pm)}(\mathbf r),
\end{equation}
which is subject to the normalization condition
\begin{equation}
\left\langle \phi_{\mathbf{k}_n}^{(\pm)} \middle| \phi_{\mathbf{k}'_n}^{(\pm)} \right\rangle = \delta^{(3)}\left(\mathbf{k}_n-\mathbf{k}'_n\right).
\label{eq:continuum_orbital_norm}
\end{equation}
The operator $\hat V_{nn}$ in Eq.~\eqref{eq:se_schroedinger2} describes the interaction of the continuum electron with the nuclei and the $(N-1)$ bound electrons of the residual ion in channel $n$.  It can be represented as:
\begin{equation}
\hat V_{nn} = \hat V_{\mathrm{nuc}} +\hat J_{nn} -\hat K_{nn} +\hat V_{\mathrm{OPP}},
\label{eq:Veff}
\end{equation}
where $\hat V_{\mathrm{nuc}}$ is the electron--nuclear attraction, whereas $\hat J_{nn}$ and $\hat K_{nn}$ are, respectively, the Coulomb and exchange interactions with the electrons of the residual ion.
The last term is an orthogonalizing pseudopotential (OPP)\cite{IVANOV20039},
\begin{equation}
\hat V_{\mathrm{OPP}} =
\lambda \sum_{i}^\mathrm{occ} |\varphi_i^{\mathrm{HF}}\rangle
\langle\varphi_i^{\mathrm{HF}}|,
\label{eq:Vopp}
\end{equation}
introduced to enforce orthogonality of the continuum orbital to the occupied canonical bound orbitals $\{\varphi_i^{\mathrm{HF}}\}$ of the Hartree--Fock reference determinant ($\lambda$ is a large positive penalty parameter).  
The remaining terms in Eq.~(\ref{eq:Veff}) are defined as follows. The nuclear attraction is given by
\begin{equation}
\hat V_{\mathrm{nuc}}(\mathbf r) = -\sum_A \frac{Z_A}{|\mathbf r-\mathbf R_A|}.
\end{equation}
The Coulomb and exchange operators can be expressed in terms of the one-particle density $\rho_{nn}$ and the one-particle density matrix
$\gamma_{nn}$ of the ionic target state $\Psi_{n}^{N-1}$ as
\begin{align}
\hat J_{nn}\phi(\mathbf r) &= \int d\mathbf r'\,\frac{\rho_{nn}(\mathbf r')}{|\mathbf r-\mathbf r'|}\phi(\mathbf r),
\label{eq:J_def}
\\
\hat K_{nn}\phi(\mathbf r)
&=\int d\mathbf r'\,\frac{\gamma_{nn}(\mathbf r,\mathbf r')}{|\mathbf r-\mathbf r'|}\phi(\mathbf r').
\label{eq:K_def}
\end{align}
The one-particle density is given by the diagonal of the one-particle density matrix,
\begin{equation}
\rho_{nn}(\mathbf r) = \gamma_{nn}(\mathbf r,\mathbf r).
\label{eq:rho_gamma}
\end{equation}
Thus, the Coulomb interaction depends only on the diagonal of $\gamma_{nn}$, whereas the nonlocal exchange interaction involves
the full one-particle density matrix.

For a correlated ionic wave function, the one-particle density matrix can be conveniently expressed in an orbital basis as
\begin{equation}
\gamma_{nn}(\mathbf r,\mathbf r') = \sum_{pq}\gamma_{pq}^{nn}\,\varphi_p^*(\mathbf r)\,\varphi_q(\mathbf r'),
\label{eq:gamma_expansion}
\end{equation}
where
\begin{equation}
\gamma_{pq}^{nn} = \left\langle \Psi_{n}^{N-1} \left| \hat a_p^\dagger \hat a_q \right| \Psi_{n}^{N-1} \right\rangle .
\label{eq:gamma_pq}
\end{equation}
Here, $\hat a_p^\dagger$ and $\hat a_q$ denote the creation and annihilation operators associated with the spin orbitals $\varphi_p$
and $\varphi_q$, respectively, and the summation runs over all the spin-orbital basis used to represent the bound-electron space.
The one-particle density-matrix elements $\gamma_{pq}^{nn}$ can be routinely evaluated for a broad range of correlated electronic-structure
methods, providing a general route for constructing the Coulomb and exchange interactions for correlated ionic states.


The Schr\"odinger equation in Eq.~\eqref{eq:se_schroedinger2} provides,
within the single-channel approximation, an exact description of the continuum orbital. Direct solution of this equation with the required asymptotic boundary conditions typically involves a partial-wave expansion of the continuum orbital and asymptotic matching of the individual
partial-wave components. For molecular systems with nonspherical, multicenter potentials, such a procedure can become cumbersome and
computationally demanding. Instead, the appropriate asymptotic boundary conditions can be incorporated directly by reformulating Eq.~\eqref{eq:se_schroedinger2} as a Lippmann--Schwinger (LS) equation~\cite{Lippmann1950, Mitroy1986,Gwmm6l57}. To this end, the effective single-channel one-electron Hamiltonian is partitioned as
\begin{equation}
-\frac{\nabla^2}{2}+\hat V_{nn} = \hat h_0+\hat V,
\label{eq:H_partition}
\end{equation}
where $\hat h_0$ is a suitably chosen zeroth-order Hamiltonian and $\hat V$
is the residual interaction. The corresponding LS equation reads
\begin{equation}
\left|\phi_{\mathbf{k}_n}^{(\pm)}\right\rangle = \left|\mathbf{k}_n^{(\pm)}\right\rangle_{\textrm{ref}} + 2\hat G_{\textrm{ref}}^{(\pm)}(k_n)\, \hat V\, \left|\phi_{\mathbf{k}_n}^{(\pm)}\right\rangle,
\label{eq:LS_basic}
\end{equation}
where the  continuum reference state satisfies
\begin{equation}
\hat h_0 \left|\mathbf{k}_n^{(\pm)}\right\rangle_{\textrm{ref}} = \frac{k_n^2}{2} \left|\mathbf{k}_n^{(\pm)}\right\rangle_{\textrm{ref}},
\label{eq:H0_state}
\end{equation}
and the Green's operator (in Helmholtz form) associated with $\hat h_0$ is
defined as
\begin{equation}
\hat G_{\textrm{ref}}^{(\pm)}(k) = \lim_{\epsilon\to0^+} \frac{1}{k^2-2\hat h_0\pm i\epsilon}.
\label{eq:G0_def}
\end{equation}
The choice of $\hat h_0$ determines both the reference continuum state $\left|\mathbf{k}_n^{(\pm)}\right\rangle_{\textrm{ref}}$ and the corresponding Green's operator.

In the present work, we consider two different partitions\cite{lucchese1986applications} of the effective one-electron Hamiltonian from Eq.~\eqref{eq:H_partition}.
In the first partition, denoted hereafter as LS-PW, the zeroth-order Hamiltonian is chosen as the kinetic-energy operator,
\begin{equation}
\hat h_0^{\mathrm{PW}} = -\frac{1}{2}\nabla^2,
\label{eq:h0_PW}
\end{equation}
and the residual interaction is therefore the full effective interaction of the continuum electron with the residual ion,
\begin{equation}
\hat V^{\mathrm{PW}} = \hat V_{nn}.
\label{eq:V_PW}
\end{equation}
The corresponding zeroth-order continuum state is a plane wave,
\begin{equation}
\langle \mathbf r|\mathbf{k}_n\rangle_{\textrm{ref}} = \frac{1}{(2\pi)^{3/2}} e^{i\mathbf{k}_n\cdot\mathbf r},
\label{eq:phi0_PW}
\end{equation}
which is an eigenstate of $\hat h_0^{\mathrm{PW}}$ with energy $k_n^2/2$.

In the second partition, denoted hereafter as LS-CW, the long-range Coulomb interaction associated with the net charge $Z_{\mathrm{net}}$ of the residual ion is included in the zeroth-order Hamiltonian,
\begin{equation}
\hat h_0^{\mathrm{CW}} = -\frac{1}{2}\nabla^2 +\hat V_{\mathrm C},
\qquad
V_{\mathrm C}(\mathbf r) = -\frac{Z_{\mathrm{net}}}{r}.
\label{eq:h0_CW}
\end{equation}
The residual interaction is then given by
\begin{equation}
\hat V^{\mathrm{CW}} = \hat V_{nn}-\hat V_{\mathrm C}.
\label{eq:V_CW}
\end{equation}
The corresponding zeroth-order state is a Coulomb wave:
\begin{widetext}
 \begin{equation}                                                                 
\langle \mathbf r|\mathbf{k}_n^{(\pm)}\rangle_{\textrm{ref}}= \frac{1}{(2\pi)^{3/2}} e^{i \mathbf{k}_n\cdot
  \mathbf{r}} \Gamma(1\pm i \eta) e^{-\frac{\pi\eta}{2}}   {_1F_1}\left(\mp i\eta ; 1; \pm ik_n r  
  -i \mathbf{k}_n\cdot\mathbf{r}  \right)
  \label{eq:CW_def}
\end{equation}  
\end{widetext}
where $\eta=-Z_{\mathrm{net}}/k_n$ is the Sommerfeld parameter and ${}_1F_1$ is the Kummer confluent hypergeometric function. This Coulomb wave is an eigenstate of $\hat h_0^{\mathrm{CW}}$ with energy $k_n^2/2$.

The two partitions also lead to different Green's operators entering the corresponding LS equations. In the LS-PW formulation, the Green's operator is the free-particle Green's operator,
\begin{equation}
\hat G_0^{(\pm)}(k_n) = \lim_{\epsilon\to0^+} \frac{1} {k_n^2+\nabla^2\pm i\epsilon},
\label{eq:G0_PW}
\end{equation}
whereas in the LS-CW formulation it is the Coulomb Green's operator,
\begin{equation}
\hat G_{\mathrm C}^{(\pm)}(k_n) = \lim_{\epsilon\to0^+} \frac{1}{k_n^2+\nabla^2-2\hat V_{\mathrm C}\pm i\epsilon}.
\label{eq:GC_def}
\end{equation}

It is useful at this point to compare the LS-PW and LS-CW formulations. Both originate from the same effective Schr\"odinger equation and differ in the partitioning of the Hamiltonian into the zeroth-order and residual interactions. 
For $Z_{\mathrm{net}}=0$, as in electron scattering from a neutral target, the Coulomb reference potential vanishes and the LS-CW formulation reduces to LS-PW. 
The distinction between the two formulations therefore becomes relevant for systems with a charged residual target. In this case, the two formulations employ different reference states and Green's operators and, consequently, different asymptotic forms of the continuum wave function. For a given partial wave, the LS-PW and LS-CW solutions can differ by an overall complex factor associated with the different reference continuum states used in the two formulations\cite{Colle1989b}. The LS-CW formulation provides the formally appropriate description of the asymptotic continuum state, since the long-range Coulomb interaction $-Z_{\mathrm{net}}/r$ is incorporated explicitly into the zeroth-order Hamiltonian. The importance of the Coulomb distortion is characterized by the Sommerfeld parameter $\eta$, whose magnitude decreases with increasing continuum-electron energy. Consequently, differences associated with the choice of  plane wave or Coulomb wave as  a reference are expected to become less pronounced at high electron energies.

These considerations are particularly relevant for practical calculations. The LS-PW formulation is more straightforward to implement and apply, whereas the use of Coulomb waves and the Coulomb Green's operator makes the LS-CW formulation computationally more involved. In addition,
for non-symmetric polyatomic systems, the LS-CW formulation requires the choice of a Coulomb center, which is not uniquely defined for a general multicenter molecular potential. This ambiguity is absent in the LS-PW formulation, where the zeroth-order Hamiltonian contains only the kinetic-energy operator. The practical and numerical consequences of these differences between the two formulations are investigated in later sections.

\subsection{Finite-basis representation in SGTOs}
The Lippmann--Schwinger equation introduced above is an operator equation. In the present work, we solve this equation in a finite auxiliary basis. To this end, we introduce a set of $M$ nonorthogonal functions
$\{\chi_\mu\}_{\mu=1}^{M}$, which defines a projector onto the corresponding
finite-dimensional subspace,
\begin{equation}
\hat P = \sum_{\mu\nu} |\chi_\mu\rangle (\mathbf S^{-1})_{\mu\nu} \langle\chi_\nu|,
\qquad
S_{\mu\nu} = \langle\chi_\mu|\chi_\nu\rangle,
\label{eq:projector_nonortho}
\end{equation}
where $\mathbf S$ is the overlap matrix of the auxiliary basis. In the present work, the auxiliary basis is composed of spherical Gaussian-type orbitals (SGTOs), which are generally defined as
\begin{equation}
\chi_\mu(\mathbf r) = N_\mu |\mathbf r-\mathbf A_\mu|^{\ell_\mu} e^{-\alpha_\mu|\mathbf r-\mathbf A_\mu|^2}
Y_{\ell_\mu m_\mu} \left(\Omega_{\mathbf r-\mathbf A_\mu}\right),
\label{eq:spherical_gaussian}
\end{equation}
where $\mathbf A_\mu$ denotes the center of the Gaussian function. The use of SGTOs is particularly convenient in the present context, as it allows the required Coulomb-type integrals to be evaluated
analytically and provides direct compatibility with standard Gaussian-based quantum-chemistry methods.

We now formulate the LS equation, Eq.~\eqref{eq:LS_basic}, in the finite auxiliary basis introduced above. 
The following derivation applies to both the LS-PW and LS-CW formulations, with $(\hat V,\hat G_{\mathrm{ref}}^{(\pm)})=(\hat V^{\mathrm{PW}},\hat G_0^{(\pm)})$ for LS-PW and $(\hat V^{\mathrm{CW}},\hat G_{\mathrm C}^{(\pm)})$ for LS-CW. The continuum orbital is expanded as
\begin{equation}
|\phi_{\mathbf k_n}^{(\pm)}\rangle = \sum_{\mu} c_{\mu}^{(\pm)}(\mathbf k_n) |\chi_\mu\rangle,
\label{eq:phi_expansion}
\end{equation}
so that the continuum problem is reduced to determining the expansion coefficients $c_{\mu}^{(\pm)}(\mathbf k_n)$. We introduce the matrix elements
\begin{align}
V_{\mu\nu} &= \langle\chi_\mu|\hat V|\chi_\nu\rangle,
\\
G_{\mu\nu}^{(\pm)} &= \left\langle\chi_\mu \left| \hat G_{\mathrm{ref}}^{(\pm)} \right| \chi_\nu \right\rangle,
\\
(\mathbf S_{\mathbf k_n}^{(\pm)})_\mu &= \left\langle \chi_\mu \middle |\mathbf{k}_n^{(\pm)}\right\rangle_{\textrm{ref}}.
\label{eq:matrix_elements}
\end{align}
Using the projector $\hat P$ of Eq.~\eqref{eq:projector_nonortho},
the LS equation is projected onto the auxiliary basis space, yielding
\begin{equation}
\mathbf c^{(\pm)}(\mathbf k_n)
=
\mathbf S^{-1}\mathbf S_{\mathbf k_n}^{(\pm)}
+
2\mathbf S^{-1}\mathbf G^{(\pm)}
\mathbf S^{-1}\mathbf V\,
\mathbf c^{(\pm)}(\mathbf k_n).
\label{eq:LS_matrix}
\end{equation}
Solving this equation for the expansion coefficients gives
\begin{equation}
\mathbf c^{(\pm)}(\mathbf k_n) = \left(\mathbf S - 2\mathbf G^{(\pm)}\mathbf S^{-1}\mathbf V \right)^{-1}
\mathbf S_{\mathbf k_n}^{(\pm)}.
\label{eq:c_coefficients}
\end{equation}
Thus, once the overlap matrix $\mathbf S$, the potential matrix $\mathbf V$, the Green's-function matrix $\mathbf G^{(\pm)}$, and the reference-state overlap vector $\mathbf S_{\mathbf k_n}^{(\pm)}$ are
known, the continuum orbital can be obtained entirely through standard matrix operations.
Equation~\eqref{eq:c_coefficients} determines the continuum orbital as the reference continuum state modified by the residual interaction, with the corresponding scattering boundary conditions incorporated through the reference Green's operator.

The specific quantities entering this finite-basis representation depend on the choice of the LS partition. In the LS-PW formulation, the reference state is a plane wave, and the overlap integrals with the auxiliary basis functions
\begin{equation}
(\mathbf S_{\mathbf k_n}^{\mathrm{PW}})_\mu = \langle \chi_\mu | \frac{1}{(2\pi)^{3/2}} e^{i\mathbf{k}_n\cdot\mathbf r}\rangle
\label{eq:PW_overlap}
\end{equation}
can be evaluated analytically for SGTOs. Similarly, the matrix elements of the free-particle Green's operator,
\begin{equation}
(G_0^{(\pm)})_{\mu\nu} =\left\langle\chi_\mu\left|\hat G_0^{(\pm)}\right|\chi_\nu\right\rangle,
\label{eq:G0_matrix}
\end{equation}
are evaluated using the analytic expressions recently developed in Ref.~\citenum{Mahato2026}. Together with the potential matrix elements $V_{\mu\nu}$, these quantities provide all matrix elements required to
solve the LS-PW equation in the SGTO basis.

The LS-CW formulation requires an additional step, since the Coulomb Green's operator must be represented in the auxiliary basis. It can be related to the free-particle Green's operator through the Dyson equation,
\begin{equation}
\hat G_{\mathrm C}^{(\pm)} = \hat G_0^{(\pm)} + 2\hat G_0^{(\pm)} \hat V_{\mathrm C} \hat G_{\mathrm C}^{(\pm)},
\label{eq:GC_Dyson_Helmholtz}
\end{equation}
where the factor of 2 arises from the Helmholtz definition of the
Green's operators used throughout this work. Projection onto the same
nonorthogonal auxiliary basis yields
\begin{equation}
\mathbf G_{\mathrm C}^{(\pm)} = \mathbf G_0^{(\pm)} + 2\mathbf G_0^{(\pm)} \mathbf S^{-1}\mathbf V_{\mathrm C}\mathbf S^{-1} \mathbf G_{\mathrm C}^{(\pm)},
\label{eq:GC_matrix_Dyson}
\end{equation}
where
\begin{equation}
(V_{\mathrm C})_{\mu\nu} = \langle\chi_\mu|\hat V_{\mathrm C}|\chi_\nu\rangle.
\label{eq:VC_matrix}
\end{equation}
Solving Eq.~\eqref{eq:GC_matrix_Dyson} for $\mathbf G_{\mathrm C}^{(\pm)}$ gives
\begin{equation}
\mathbf G_{\mathrm C}^{(\pm)} = \left[ \mathbf I - 2\mathbf G_0^{(\pm)} \mathbf S^{-1}\mathbf V_{\mathrm C}\mathbf S^{-1} \right]^{-1} \mathbf G_0^{(\pm)},
\label{eq:GC_matrix_solution}
\end{equation}
where $\mathbf I$ denotes the identity matrix.

In addition, the LS-CW formulation requires the overlaps between the auxiliary SGTO basis functions and the Coulomb-wave reference state defined in Eq.~\eqref{eq:CW_def},
\begin{equation}
(\mathbf S_{\mathbf k_n}^{\mathrm{CW}(\pm)})_\mu
=\left\langle
\chi_\mu
\middle|
\mathbf k_n^{(\pm)}
\right\rangle_{\mathrm{ref}}.
\label{eq:Coulomb_overlap}
\end{equation}
These overlaps can be evaluated in different ways. One possibility is to fit the Coulomb wave into an auxiliary expansion using London-type orbitals as it was done in Ref.~\citenum{Skomorowski2021}, after which
the required overlaps can be evaluated analytically. Alternatively, the three-dimensional integral in Eq.~\eqref{eq:Coulomb_overlap} can be evaluated directly by numerical quadrature. Since the corresponding
integrand is nonsingular, the latter approach permits the overlaps to be evaluated numerically to high accuracy.

In summary, Eq.~\eqref{eq:c_coefficients} provides the central working equation for determining the continuum-orbital expansion coefficients in the finite auxiliary SGTO basis. The continuum problem is thereby reduced to a sequence of standard matrix operations. Although the continuum orbital itself is represented in a finite $L^2$ basis, the appropriate outgoing or incoming boundary conditions are incorporated through the reference state and the corresponding Green's operator. The LS-PW and LS-CW formulations can thus be treated within the same finite-basis framework, differing only in the reference-state overlaps and Green's-function matrix elements entering Eq.~\eqref{eq:c_coefficients}.

\subsection{Inclusion of interchannel coupling}
The formulation presented so far has been restricted to the single-channel approximation, in which coupling between continuum states associated with different ionic channels is neglected. We now generalize the formalism to include interchannel coupling within a close-coupling expansion. We consider $N_{\mathrm c}$ coupled open channels. The effects of closed channels are not treated explicitly here, but may in principle be incorporated through an inclusion of additional optical potential.

For a specified asymptotic channel $n$, the single-channel ansatz of
Eq.~\eqref{eq:2} is generalized to
\begin{equation}
\Psi_{\mathbf k_n}^{(\pm), N}
(\mathbf r_1,\ldots,\mathbf r_{N}) = \sum_{m=1}^{N_{\mathrm c}}
\hat{\mathcal A} \left\{ \phi_{n,\mathbf k_m}^{(\pm)}(\mathbf r_1)\, \Psi_m^{N-1}(\mathbf r_2,\ldots,\mathbf r_{N}) \right\}.
\label{eq:2cc}
\end{equation}
Here, $\phi_{n,\mathbf k_m}^{(\pm)}$ denotes the continuum component
in channel $m$, with the corresponding momentum $\mathbf k_m$, of the
solution associated asymptotically with channel $n$.
Interchannel coupling generates contributions in all coupled channels,
whereas the reference continuum state is present only in the selected
asymptotic channel $n$. Accordingly, the reference channel vector can
be written as
\begin{equation}
\left|\boldsymbol{\Phi}_{n,\mathrm{ref}}^{(\pm)}\right\rangle =
\begin{pmatrix}
0 \\
\vdots \\
\left|\mathbf{k}_n^{(\pm)}\right\rangle_{\mathrm{ref}} \\
\vdots \\
0
\end{pmatrix},
\label{eq:reference_channel_vector}
\end{equation}
where the only nonzero component corresponds to channel $n$.

The coupled-channel Lippmann--Schwinger equation can then be written
compactly in channel space as
\begin{equation}
\left|\boldsymbol{\Phi}_n^{(\pm)}\right\rangle =
\left|\boldsymbol{\Phi}_{n,\mathrm{ref}}^{(\pm)}\right\rangle +
2\hat{\mathbf G}_{\mathrm{ref}}^{(\pm)}
\hat{\mathbf V}
\left|\boldsymbol{\Phi}_n^{(\pm)}\right\rangle,
\label{eq:LS_multichannel}
\end{equation}
where
\begin{equation}
\left|\boldsymbol{\Phi}_n^{(\pm)}\right\rangle =
\begin{pmatrix}
\left|\phi_{n,\mathbf k_1}^{(\pm)}\right\rangle \\
\left|\phi_{n,\mathbf k_2}^{(\pm)}\right\rangle \\
\vdots \\
\left|\phi_{n,\mathbf k_{N_{\mathrm c}}}^{(\pm)}\right\rangle
\end{pmatrix}.
\label{eq:channel_vector}
\end{equation}
The Green's operator is diagonal in channel space,
\begin{equation}
\hat{\mathbf G}_{\mathrm{ref}}^{(\pm)} =
\begin{pmatrix}
\hat G_{\mathrm{ref},1}^{(\pm)} & 0 & \cdots & 0 \\
0 & \hat G_{\mathrm{ref},2}^{(\pm)} & \cdots & 0 \\
\vdots & \vdots & \ddots & \vdots \\
0 & 0 & \cdots & \hat G_{\mathrm{ref},N_{\mathrm c}}^{(\pm)}
\end{pmatrix},
\label{eq:G_channel}
\end{equation}
with each diagonal block evaluated at the continuum energy
$\varepsilon_m=k_m^2/2$ of the corresponding channel.
In contrast, the interaction operator has the general block structure
\begin{equation}
\hat{\mathbf V}
=
\begin{pmatrix}
\hat V_{11} & \hat V_{12} & \cdots & \hat V_{1N_{\mathrm c}} \\
\hat V_{21} & \hat V_{22} & \cdots & \hat V_{2N_{\mathrm c}} \\
\vdots & \vdots & \ddots & \vdots \\
\hat V_{N_{\mathrm c}1} & \hat V_{N_{\mathrm c}2} & \cdots &
\hat V_{N_{\mathrm c}N_{\mathrm c}}
\end{pmatrix}.
\label{eq:V_channel}
\end{equation}
The diagonal blocks $\hat V_{mm}$ correspond to the intrachannel
interactions introduced in the single-channel model, whereas the
off-diagonal blocks $\hat V_{mm'}$ describe coupling between different
ionic channels. The off-diagonal Coulomb and exchange contributions can
be constructed in direct analogy with their diagonal counterparts, but
using the one-particle transition-state density matrix between the ionic target
states $m$ and $m'$,
\begin{equation}
\gamma_{pq}^{mm'}
=
\left\langle
\Psi_m^{N-1}
\left|
\hat a_p^\dagger \hat a_q
\right|
\Psi_{m'}^{N-1}
\right\rangle.
\label{eq:gamma_pq_mmprime}
\end{equation}
The corresponding transition-state density matrix in spin-orbital representation is
\begin{equation}
\gamma_{mm'}(\mathbf r,\mathbf r') = \sum_{pq} \gamma_{pq}^{mm'}\,\varphi_p^*(\mathbf r)\, \varphi_q(\mathbf r'),
\label{eq:gamma_nm}
\end{equation}
with the transition-state density
\begin{equation}
\rho_{mm'}(\mathbf r) = \gamma_{mm'}(\mathbf r,\mathbf r).
\label{eq:rho_nm}
\end{equation}
The off-diagonal Coulomb and exchange operators are consequently given by
\begin{align}
\hat J_{mm'}\phi(\mathbf r) &=
\int d\mathbf r'\,
\frac{\rho_{mm'}(\mathbf r')}
     {|\mathbf r-\mathbf r'|}
\phi(\mathbf r),
\label{eq:J_offdiag}
\\
\hat K_{mm'}\phi(\mathbf r)
&=
\int d\mathbf r'\,
\frac{\gamma_{mm'}(\mathbf r,\mathbf r')}
     {|\mathbf r-\mathbf r'|}
\phi(\mathbf r').
\label{eq:K_offdiag}
\end{align}
Thus, the interchannel coupling is determined directly by the one-particle transition density matrices between the corresponding ionic states. As in the case of the state one-particle density matrices introduced above, the transition-state density matrices $\gamma_{pq}^{mm'}$ are standard quantities that can be routinely evaluated for a broad range of correlated electronic wave functions.

The finite-basis formulation introduced in Sec.~IIB extends naturally to the close-coupling case. Each channel component is expanded in the same auxiliary SGTO basis,

\begin{equation}
\left|\phi_{n,\mathbf k_m}^{(\pm)}\right\rangle =
\sum_{\mu=1}^{M} c_{\mu n}^{(\pm)}(\mathbf k_m) |\chi_\mu\rangle,
\label{eq:phi_cc_expansion}
\end{equation}

and the coefficients for all channels are collected into a single vector of dimension $N_{\mathrm c}\times M$,
\begin{equation}
\mathbf C_n^{(\pm)}
=
\begin{pmatrix}
\mathbf c_n^{(\pm)}(\mathbf k_1) \\
\mathbf c_n^{(\pm)}(\mathbf k_2) \\
\vdots \\
\mathbf c_n^{(\pm)}(\mathbf k_{N_{\mathrm c}})
\end{pmatrix}.
\label{eq:C_channel_vector}
\end{equation}
The overlap and Green's-function matrices are block diagonal in the
combined channel--auxiliary-basis space,
\begin{equation}
\boldsymbol{\mathcal S}
=
\operatorname{diag}(\mathbf S,\ldots,\mathbf S),
\qquad
\boldsymbol{\mathcal G}^{(\pm)}
=
\operatorname{diag}
\left(
\mathbf G_1^{(\pm)},\ldots,
\mathbf G_{N_{\mathrm c}}^{(\pm)}
\right),
\label{eq:SG_block}
\end{equation}

whereas the interaction matrix contains both diagonal and off-diagonal channel blocks,
\begin{widetext}
\begin{equation}
\boldsymbol{\mathcal V}
=
\begin{pmatrix}
\mathbf V_{11} & \mathbf V_{12} & \cdots & \mathbf V_{1N_{\mathrm c}} \\
\mathbf V_{21} & \mathbf V_{22} & \cdots & \mathbf V_{2N_{\mathrm c}} \\
\vdots & \vdots & \ddots & \vdots \\
\mathbf V_{N_{\mathrm c}1} & \mathbf V_{N_{\mathrm c}2} & \cdots &
\mathbf V_{N_{\mathrm c}N_{\mathrm c}}
\end{pmatrix},
\qquad
(V_{mm'})_{\mu\nu}
=
\langle\chi_\mu|\hat V_{mm'}|\chi_\nu\rangle.
\label{eq:V_block}
\end{equation}
\end{widetext}
The reference-state overlap vector has the corresponding block form

\begin{equation}
\boldsymbol{\mathcal S}_{\mathbf k_n}^{(\pm)}
=
\begin{pmatrix}
\mathbf 0 \\
\vdots \\
\mathbf S_{\mathbf k_n}^{(\pm)} \\
\vdots \\
\mathbf 0
\end{pmatrix},
\label{eq:Sk_channel}
\end{equation}
where only the block corresponding to the selected asymptotic channel $n$ is nonzero.

With these definitions, the close-coupling equations retain the same algebraic structure as in the single-channel case. In particular, the block-matrix generalization of Eq.~\eqref{eq:c_coefficients} gives the complete vector of continuum expansion coefficients as
\begin{equation}
\mathbf C_n^{(\pm)}
=
\left[
\boldsymbol{\mathcal S}
-
2\boldsymbol{\mathcal G}^{(\pm)}
\boldsymbol{\mathcal S}^{-1}
\boldsymbol{\mathcal V}
\right]^{-1}
\boldsymbol{\mathcal S}_{\mathbf k_n}^{(\pm)}.
\label{eq:C_multichannel}
\end{equation}
Since $\boldsymbol{\mathcal S}_{\mathbf k_n}^{(\pm)}$ has only one nonzero block, corresponding to the selected asymptotic channel $n$, the continuum coefficients in an individual channel $m$ can equivalently be written as
\begin{equation}
\mathbf c_n^{(\pm)}(\mathbf k_m)
=
\left\{
\left[
\boldsymbol{\mathcal S}
-
2\boldsymbol{\mathcal G}^{(\pm)}
\boldsymbol{\mathcal S}^{-1}
\boldsymbol{\mathcal V}
\right]^{-1}
\right\}_{mn}
\mathbf S_{\mathbf k_n}^{(\pm)},
\label{eq:c_multichannel_block}
\end{equation}
where $\{\cdots\}_{mn}$ denotes the $M\times M$ block connecting the selected asymptotic channel $n$ to an open channel $m$.

Thus, the inclusion of interchannel coupling does not alter the basic structure of the finite-basis LS solution. The single-channel vectors and matrices are simply promoted to block vectors and matrices in the combined channel--auxiliary-basis space, with the off-diagonal blocks of the interaction matrix accounting for coupling between different ionic channels. The close-coupling problem can therefore be solved using the
same matrix operations as in the single-channel case, with the dimension of the matrices increased from $M$ to $N_{\mathrm c}\times M$.  In the absence of interchannel coupling, the off-diagonal blocks of $\boldsymbol{\mathcal V}$ vanish, and the channel components decouple.
Consequently, Eq.~\eqref{eq:c_multichannel_block} reduces to
\begin{equation}
\mathbf c_n^{(\pm)}(\mathbf k_m) =
\delta_{mn}\, \mathbf c^{(\pm)}(\mathbf k_n),
\label{eq:cc_single_limit}
\end{equation}
thereby recovering the single-channel formulation derived in Sec.~IIB.


\subsection{Auger decay widths calculations}
The continuum orbital obtained within the LS framework described above can be employed to model a broad range of processes involving electron emission. As a first application, in the present work we use it to calculate normal Auger decay spectra for a set of benchmark atoms and small molecules. This constitutes a direct extension of previous EOM-CCSD treatment of Auger decay, in which the bound many-electron states were described at the same level of theory, while the emitted electron was represented by a model continuum orbital, either a plane wave or a Coulomb wave with an effective charge.~\cite{Skomorowski2021, Skomorowski2021_2} Since the many-body formulation of the Auger decay widths remains unchanged, we recall here only its essential elements and focus on the new continuum description.

We consider Auger decay as the second step of a two-step process, in which core ionization is followed by electronic relaxation and emission of an Auger electron. For normal Auger decay originating from a neutral
closed-shell system, the process can be schematically represented as
\begin{equation}
\Psi_I^N \;\longrightarrow\; \Psi_n^{N-1} + e^-,
\label{eq:normal_auger}
\end{equation}
where $\Psi_I^N$ denotes the initial core-ionized state and $\Psi_n^{N-1}$ the final dicationic state defining channel $n$. The core-ionized state is treated here as a metastable electronic state
embedded in the ionization continuum. Within the Fano--Feshbach formalism, its partial decay width into channel $n$ can be expressed in Fermi-golden-rule form as
\begin{equation}
\Gamma_{I \to n}
=
2\pi k_n 
\int d \Omega_{k_n}\,
\left|
\left\langle
\Psi_I^{N}
\left|
\hat H^{N}-E_I
\right|
\Psi_{\mathbf k_n}^{(-),N}
\right\rangle
\right|^2 ,
\label{eq:auger_width_general}
\end{equation}
where $\Psi_{\mathbf k_n}^{(-),N}$ denotes the $N$-electron continuum state associated with channel $n$, consisting asymptotically of the dicationic state $\Psi_n^{N-1}$ and an emitted Auger electron with
momentum $\mathbf k_n$. The continuum state $\Psi_{\mathbf k_n}^{(-),N}$ can be constructed using either the
single-channel approximation or the multichannel close-coupling expansion introduced above -- Eq.~\eqref{eq:auger_width_general} applies to both formulations.
The integration is performed over all emission directions
$\Omega_{k_n} = (\theta_{k_n}, \phi_{k_n})$ at the electron energy fixed by energy conservation,
\begin{equation}
\varepsilon_n =
\frac{k_n^2}{2} = E_I-E_n,
\label{eq:auger_energy_conservation}
\end{equation}
where $E_n$ is the energy of the final dicationic state $\Psi_n^{N-1}$. 
The $(-)$ scattering solution is used to represent the final continuum state of the autoionization process.%

Application of the isolated-resonance expression from Eq.~\eqref{eq:auger_width_general} requires a strict separation of the initial core-ionized state from the continuum sector, which can be formally achieved through Feshbach projection. For Auger decay, however, this separation can be conveniently realized within the core--valence separation (CVS) framework.
In the present work, the initial core-ionized states $\Psi_I^{N}$ and the dicationic states $\Psi_{n}^{N-1}$ defining the final Auger channels are described at the frozen-core-CVS-EOM-IP-CCSD and EOM-DIP-CCSD levels of theory, respectively, following the formulation used in Refs.~\citenum{Skomorowski2021, Skomorowski2021_2}.

Assuming that the continuum orbital is orthogonal to the bound orbital space and employing the frozen-core approximation, the contribution to the Auger decay amplitude can be expressed in terms of the electron--electron interaction and the two-body Dyson transition amplitudes. The partial width for channel \(n\) in the single-channel approximation takes the form
\begin{widetext}
\begin{equation}
\Gamma_{I \to n}
= 2\pi k_n g_{\alpha(n)} \int d \Omega_{k_n} \, \left( \frac{1}{2} \sum_{pqr} \langle pq\Vert\mathbf k_n r\rangle\, R_{I n;pqr}\right)
\left( \frac{1}{2} \sum_{pqr} \langle\mathbf k_n r\Vert pq\rangle\, R_{n I;rpq}
\right),
\label{eq:auger_width_dyson}
\end{equation}
\end{widetext}
where $\langle pq\Vert\mathbf k_n r\rangle$ are antisymmetrized mixed
bound--continuum two-electron integrals. The indices $p$, $q$, and $r$ refer to bound molecular spin orbitals, whereas $\mathbf k_n \equiv \phi_{\mathbf{k}_n}^{(-)}$  denotes
the continuum orbital associated with the emitted Auger electron. 
The factor $g_{\alpha(n)}$ accounts for the spin degeneracy of channel
$n$, assuming that the matrix elements are evaluated for an $\alpha$-spin continuum electron.~\cite{Skomorowski2021}. 
The two-body Dyson transition amplitudes entering
Eq.~\eqref{eq:auger_width_dyson} are defined as
\begin{widetext}
\begin{equation}
R_{I n;pqr} = \left\langle \Psi_I^{N} \left| \hat a_p^\dagger \hat a_q^\dagger \hat a_r \right| \Psi_{n}^{N-1} \right\rangle,
\qquad
R_{n I;rpq} = \left\langle \Psi_{n}^{N-1} \left| \hat a_r^\dagger \hat a_p \hat a_q \right| \Psi_I^{N} \right\rangle .
\label{eq:two_body_dyson}
\end{equation}
\end{widetext}
These quantities contain the correlated many-electron information associated with the initial and final bound-domain states and can be evaluated directly from the corresponding EOM-CCSD wave functions\cite{Skomorowski2021}. 
We emphasize that the simple form of the above expression for $\Gamma_{I \to n}$ results from the use of a common set of canonical molecular orbitals ${\varphi_p}$ within the EOM-CC framework to describe both the initial $\Psi_I^N$ and final bound states $\Psi_n^{N-1}$. In addition, within the frozen-core CVS approximation, the corresponding one-body Dyson orbital vanishes identically, leaving only the two-electron coupling contribution to the partial decay width.

Equation~\eqref{eq:auger_width_dyson} provides a particularly useful separation between the bound-state and continuum parts of the calculation. The correlated many-electron description of the decay is contained in the two-body Dyson amplitudes, whereas the continuum
electron enters through the mixed two-electron integrals $\langle pq\Vert\mathbf k_n r\rangle$. Consequently, once the continuum
orbital $\phi_{\mathbf k_n}^{(-)}$ has been determined, the partial Auger width is obtained by contracting the corresponding mixed
bound--continuum integrals with the two-body Dyson amplitudes and integrating over the emission direction.

When interchannel coupling is included,
the continuum state associated asymptotically with channel $n$ contains
contributions from all coupled channels $m$, and the corresponding
partial width of Eq.~\eqref{eq:auger_width_dyson} becomes
\begin{widetext}
\begin{equation}
\Gamma_{I \to n}^{\mathrm{IC}}
= 2\pi k_n \int d \Omega_{k_n}\, \left( \frac{1}{2} \sum_{m=1}^{N_c} g_{\alpha(m)}^{1/2}\sum_{pqr} \langle pq\Vert\mathbf k_{n}^m r\rangle\, R_{I m;pqr}\right)
\left( \frac{1}{2} \sum_{m=1}^{N_c} g_{\alpha(m)}^{1/2}\sum_{pqr} \langle\mathbf k_{n}^m r\Vert pq\rangle\, R_{m I;rpq}
\right),
\label{eq:auger_width_dyson_cc}
\end{equation}
\end{widetext}
where $\mathbf k_{n}^m \equiv\phi_{n,\mathbf k_m}^{(-)}$ denotes the appropriate continuum component.
Thus, the contributions from the different channel components enter
coherently at the amplitude level. In the absence of interchannel
coupling, only the $m=n$ component remains, and
Eq.~\eqref{eq:auger_width_dyson_cc} reduces to the single-channel
expression of Eq.~\eqref{eq:auger_width_dyson}.

The essential improvement of the present approach over the treatment employed in Ref.~\citenum{Skomorowski2021} therefore lies in the representation of the continuum
orbital. Here, $\phi_{\mathbf k_n}^{(-)}$ is obtained by solving the Lippmann--Schwinger equation in the auxiliary SGTO basis introduced above, using either the LS-PW or LS-CW formulation. In the single-channel approximation, the resulting continuum orbital is used directly in the mixed two-electron integrals entering Eq.~\eqref{eq:auger_width_dyson}. When interchannel coupling is included, the corresponding channel component $\phi_{n,\mathbf k_m}^{(-)}$  obtained from the close-coupling solution of Eq.~\eqref{eq:C_multichannel} is used instead.  Thus, the bound-state EOM-CCSD description and the associated two-body Dyson amplitudes are retained throughout, while the treatment of the continuum is systematically extended from the model orbitals employed previously to the LS-PW and LS-CW formulations, including their multichannel generalizations. The impact of these different continuum treatments on the calculated Auger decay widths is examined in the following section.

\section{Numerical illustration}

\subsection{Computational details}

The numerical implementation of the continuum representation introduced above is assessed for normal Auger decay in several benchmark systems. The calculations are carried out in two stages. In the first stage, the
bound electronic states and the corresponding two-body Dyson amplitudes are computed following the procedure of Ref.~\citenum{Skomorowski2021, Skomorowski2021_2}. The initial core-ionized states are described at the frozen-core CVS-EOM-IP-CCSD level, whereas the final dicationic states defining the Auger decay channels
are described using EOM-DIP-CCSD. Standard correlation-consistent Dunning's basis sets are employed for the bound-state calculations.

In the second stage, the continuum orbitals are determined by solving the Lippmann--Schwinger equation using the LS-PW or LS-CW formulation described above. The resulting continuum orbitals are combined with the
two-body Dyson amplitudes through the mixed bound--continuum two-electron integrals to obtain the Auger decay widths.

The continuum orbitals are represented in a separate auxiliary basis of SGTOs. For each angular momentum included in the expansion, we employ an even-tempered sequence of Gaussian exponents,
\begin{equation}
\alpha_i = \alpha_{\max}\beta^{-i}, \qquad i=0,\ldots,N-1,
\label{eq:continuum_exponents}
\end{equation}
where $N$ is the number of radial functions for each angular momentum,
$\alpha_{\max}$ is the largest exponent, and $\beta$ is the scaling
factor, typically chosen in the range $1.6$--$2.0$. In the present work, the same sequence of exponents is used for all angular momenta. In the following, such a continuum basis is denoted by
\begin{equation}
[N(spdf\ldots);\beta,\alpha_{\max}],
\label{eq:continuum_basis_notation}
\end{equation}
where the symbols in parentheses specify the angular momenta included in the basis. Thus, for example, $[23(spdf);1.6,2000]$ denotes a basis containing 23 functions for each of the $s$, $p$, $d$, and $f$ angular
momenta, with exponents generated according to Eq.~\eqref{eq:continuum_exponents} using $\beta=1.6$ and
$\alpha_{\max}=2000$. 

This very simple construction of the continuum basis has several practical advantages. No optimization of individual exponents is required, and basis-set convergence can be assessed systematically by increasing the number of functions and the maximum angular momentum. Moreover, the geometric spacing of the exponents provides a balanced coverage of different radial length scales while reducing the occurrence of nearly redundant functions with very similar exponents. Such an even-tempered distribution is also well matched to the behavior of the Green's-function matrix elements discussed in Ref.~\citenum{Mahato2026}.

The angular integration over the direction of the continuum momentum $\Omega_k$ entering the Auger decay widths is performed using Lebedev quadrature, with an integration order of $N_{\mathrm{int}}=5$. The molecular geometries are the same as those employed in Ref.~\citenum{Skomorowski2021_2}.
The full implementation and all calculations were carried out using a developer's version of the Q-Chem program package.

\subsection{K-shell Auger decay in Ne}

As a first benchmark of the present continuum representation for Auger decay, we consider K-shell-ionized neon, Ne$^{+}(1s^{-1})$. Neon provides a particularly suitable reference system due to its well-resolved atomic
Auger spectrum, the absence of vibrational broadening, and the unambiguous assignment of the dominant transitions\cite{Albiez1990Z, Muller2017, Howat1978, Tulkki1992, Levasalmi1992, Coreno1999}. Therefore, it has been widely used for benchmarking theoretical descriptions of Auger decay. The Ne K--LL Auger spectrum is dominated by five channels leading to the residual ionic states $^{1}$D$(2p^{-2})$, $^{1}$S$(2p^{-2})$,
$^{3}$P$(2s^{-1}2p^{-1})$, $^{1}$P$(2s^{-1}2p^{-1})$, and $^{1}$S$(2s^{-2})$. Their branching ratios have been determined in several experiments, with good agreement among the reported values\cite{Albiez1990Z}. For comparison with the calculated total Auger width, we consider exclusively one-electron Auger decay, for which the estimated width of $242 \pm 10$~meV is derived from the relative photoionization cross sections and resonance strengths reported in Table~5/6 of Ref.~\citenum{Muller2017}.

We begin the discussion of the numerical results with Table~\ref{tab:Ne_ccpvqz_1p6}, which presents the partial Auger widths obtained using the cc-pVQZ basis set for the bound-state calculations and the $[29(spd);1.6, 13000]$ auxiliary SGTO basis for the continuum.
Throughout this work, we use the notation B/C to denote the bound-state (B) and continuum SGTO (C) basis sets, respectively. The results in Table~\ref{tab:Ne_ccpvqz_1p6} are converged with respect to the size of the continuum basis. The columns correspond to different treatments of the continuum orbital: the exact plane wave (PW), the plane wave projected onto the auxiliary SGTO basis (Proj-PW), the Lippmann--Schwinger solutions using plane-wave (LS-PW) and Coulomb-wave (LS-CW) reference states, and their multichannel variants with interchannel coupling, denoted as LS-PW-IC and LS-CW-IC, respectively. Due to the spherical symmetry of the atom, only a single angular momentum $l$ is required in the continuum SGTO basis for each single-channel calculation, whereas all relevant angular momenta ($s$, $p$, and $d$) are included when interchannel coupling is considered.

The Proj-PW calculation provides a direct test of the finite SGTO
representation of the continuum. Here, the interaction in the LS
equation is neglected, such that the projected solution corresponds to
the representation of the plane wave in the auxiliary basis. The PW and
Proj-PW widths are identical to the reported precision for all channels,
confirming that the chosen SGTO basis provides an essentially complete
representation of the plane wave for the quantities considered here.

\begin{table*}
\scriptsize
\caption{
Auger electron energies (in eV) and  partial decay widths $\Gamma_i$ (in meV) for the Ne K--LL
Auger transitions, calculated using EOM-CCSD with different representations of the continuum orbital and the cc-pVQZ / $[29(spd);1.6, 13000]$ basis sets.
}
\label{tab:Ne_ccpvqz_1p6}
\begin{tabular}{lcrrrrrrr} 
\hline
\multirow{2}{*}{Channel} & \multirow{2}{*}{Auger energy} & \multicolumn{6}{c}{Continuum model} &  \multirow{2}{*}{Exp.\cite{Albiez1990Z, Muller2017}} \\
\cline{3-8}
&
& PW  
& Proj-PW   
& LS-PW   
& LS-CW 
& LS-PW-IC
& LS-CW-IC
\\
\hline
$^1$D$(2p^{-2})$
& 803.6
& 155.5 & 155.5 & 131.2 & 140.5 & 138.5 & 147.4
& 147$\pm$6
\\
$^1$S$(2p^{-2})$
& 800.1
& 4.5 & 4.5 & 17.4 & 17.4 & 21.5 & 22.3
& 23$\pm$1
\\
$^1$P$(2s^{-1}2p^{-1})$ 
& 770.5
& 5.1 & 5.1 & 49.8 & 48.3 & 41.9 & 40.7
& 42$\pm$2
\\
$^1$S$(2s^{-2})$
& 746.6
& 18.7 & 18.7 & 18.4 & 17.4 & 14.5 & 14.4
& 15$\pm$1
\\
$^3$P$(2s^{-1}2p^{-1})$
& 781.2
& 115.9 & 115.9 & 16.0 & 15.4 & 15.9 & 15.4
& 15$\pm$1
\\
\hline
$\sum \Gamma_i$    
& 
& 299.8 & 299.8 & 232.8 & 239.0 & 232.3 & 240.2
& 242 $\pm$ 10\\  
\hline                  
\end{tabular} 
\end{table*}

As it was discussed previously~\cite{Skomorowski2021_2}, the simple PW model gives a very poor description of the
experimental branching pattern. In particular, the relative strengths
of the $^1$P$(2s^{-1}2p^{-1})$ and $^3$P$(2s^{-1}2p^{-1})$ channels are
qualitatively incorrect, with the $^3$P channel strongly dominating,
while the $^1$S$(2p^{-2})$ channel is severely underestimated. Including
the electron--ion interaction through the LS-PW formulation largely
corrects these deficiencies, yielding 49.8 and 16.0~meV for the $^1$P
and $^3$P channels, respectively, and increasing the $^1$S$(2p^{-2})$
width from 4.5 to 17.4~meV. The total width is simultaneously reduced
from 299.8 to 232.8~meV, close to the experimental estimate of
$242\pm10$~meV.

The LS-CW formulation leads to a further, but considerably smaller,
change in the calculated widths. The most pronounced effect is observed
for the dominant $^1$D$(2p^{-2})$ channel, whereas the remaining partial
widths differ from the LS-PW results by only a few percent. This
relatively small difference indicates that, at the high electron
energies characteristic of Ne K--LL Auger decay, the inclusion of the
electron--ion interaction is substantially more important than the
choice of the asymptotic reference state.

Finally, interchannel coupling leads to an additional redistribution of
the decay intensity. Its effect is particularly important for the two
$^1$S channels, whose LS-CW widths change from 17.4 and 17.4~meV to
22.3 and 14.4~meV, respectively, in close agreement with experiment.
The dominant $^1$D$(2p^{-2})$ width is also improved from 140.5 to
147.4~meV. Consequently, the LS-CW-IC calculation provides very good
agreement with both the individual experimental partial widths and the
total one-electron Auger width. In summary, Table~\ref{tab:Ne_ccpvqz_1p6} demonstrates a clear and systematic improvement of the calculated
partial widths as the electron--ion interaction and interchannel
coupling are incorporated into the continuum treatment. 

The results from Table~\ref{tab:Ne_ccpvqz_1p6}  were obtained without the orthogonalizing pseudopotential defined in Eq.~\eqref{eq:Vopp}. Its inclusion has a negligible effect, changing the partial width of any individual channel by no more than 0.2~meV, as shown in Table~S1 of the Supporting Information.

\begin{figure}
    \centering
    \includegraphics[width=0.90\linewidth]{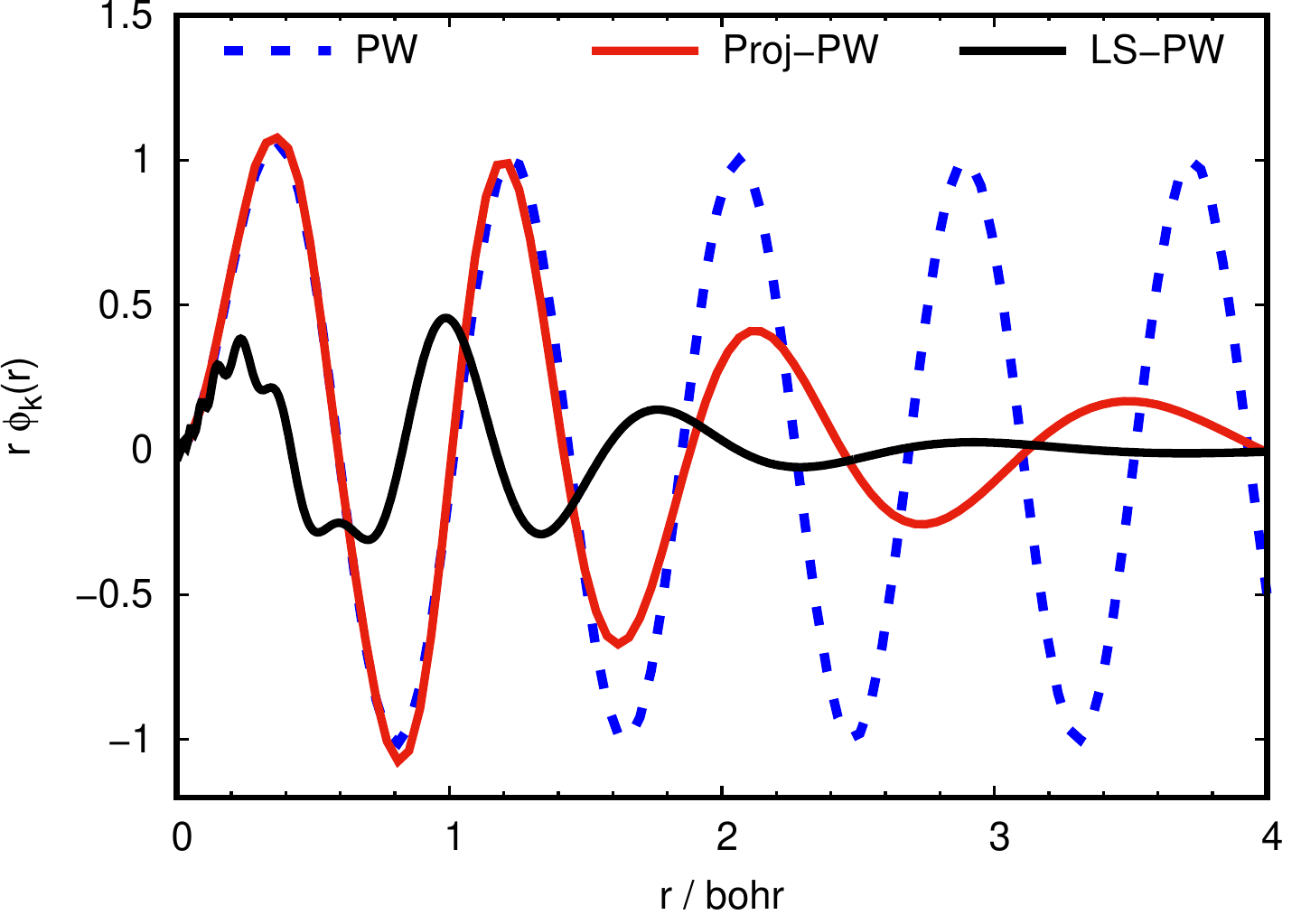}
    \caption{ Radial part of the continuum orbital for the $^3$P channel (multiplied by $r$) obtained using the exact plane wave (PW), its projection onto the SGTO basis (Proj-PW), and the LS-PW formulation. The $[29(p);1.6,13000]$ continuum SGTO basis is used for the Proj-PW and LS-PW curves.
    }
    \label{fig:cont_orb}
\end{figure}

To illustrate how the continuum orbital changes upon inclusion of the electron--ion interaction, Fig.~\ref{fig:cont_orb} shows its radial part for the $^3$P channel. In the absence of interchannel coupling, symmetry restricts the continuum electron in this channel to a single $p$ partial wave. The exact plane wave (PW) is compared with its projection onto the finite SGTO basis (Proj-PW) and with the corresponding LS-PW solution. The exact plane wave exhibits oscillatory behavior over the entire radial range. Its projection onto the SGTO basis reproduces only the first few oscillations and is subsequently damped by the Gaussian asymptotics of the basis functions. Nevertheless, as shown in Table~\ref{tab:Ne_ccpvqz_1p6}, the PW and Proj-PW models yield identical partial widths of 115.9~meV. This demonstrates that, despite the very different long-range behavior of the two continuum orbitals, the finite SGTO representation retains the part of the continuum wave function relevant to the Auger transition matrix element.

A much more pronounced change occurs when the electron--ion interaction is included through the LS-PW formulation. Relative to Proj-PW, the nodal structure of the continuum orbital is shifted and its amplitude in the interaction region is substantially modified. These changes reduce the calculated $^3$P partial width from 115.9 to 16.0~meV, bringing it into close agreement with experiment. Figure~\ref{fig:cont_orb} therefore illustrates an important feature of the present approach: although the finite SGTO expansion does not reproduce the oscillatory long-range tail of the continuum orbital, it accurately represents the short-range region that governs the bound--continuum coupling and, consequently, the Auger decay width. 

\begin{table}
\scriptsize
\caption{
Effect of the bound-state basis set on the partial and total Auger decay widths (meV) for Ne K--LL transitions, calculated using the LS-CW-IC model with the $[29(spd);1.6,13000]$ SGTO continuum basis.
}
\label{tab:Ne_bound_bs}
\begin{tabular}{lrrrr} 
\hline
\multirow{2}{*}{Channel} & \multicolumn{3}{c}{Bound-state basis set} &  \multirow{2}{*}{Exp.\cite{Albiez1990Z, Muller2017}} \\
\cline{2-4}
& cc-pVTZ 
& aug-cc-pVTZ
& cc-pVQZ
\\
\hline
$^1$D$(2p^{-2})$
& 143.8 & 142.9 & 147.4
& 147$\pm$6
\\
$^1$S$(2p^{-2})$
& 21.5 & 21.4 & 22.3
& 23$\pm$1
\\
$^1$P$(2s^{-1}2p^{-1})$ 
& 38.7 & 38.5 & 40.7
& 42$\pm$2
\\
$^1$S$(2s^{-2})$
& 14.9 & 14.9 & 14.4
& 15$\pm$1
\\
$^3$P$(2s^{-1}2p^{-1})$
& 14.6 & 14.5 & 15.4
& 15$\pm$1
\\
\hline
$\sum \Gamma_i$    
& 233.5 & 232.2 & 240.2
& 242 $\pm$ 10\\  
\hline                  
\end{tabular} 
\end{table}

We next examine separately the dependence of the calculated widths on the bound-state and continuum basis sets. We first consider the bound-state basis, with the LS-CW-IC results obtained using cc-pVTZ, aug-cc-pVTZ, and cc-pVQZ reported in Table~\ref{tab:Ne_bound_bs}. The dependence on the bound-state basis is generally weak, and the calculated partial and total widths remain within the experimental uncertainties for all three basis sets. In particular, augmentation of the cc-pVTZ basis has a negligible effect, while the differences between the cc-pVTZ and cc-pVQZ partial widths remain below 5\% for all channels.
These results highlight an important feature of the present formulation, in which the bound-state and continuum representations are treated independently. Convergence with respect to the two basis sets can therefore be controlled separately, and diffuse functions are not required in the bound-domain basis to describe the continuum electron. 

\begin{figure}[t!]
    \centering
    \includegraphics[width=1.0\linewidth]{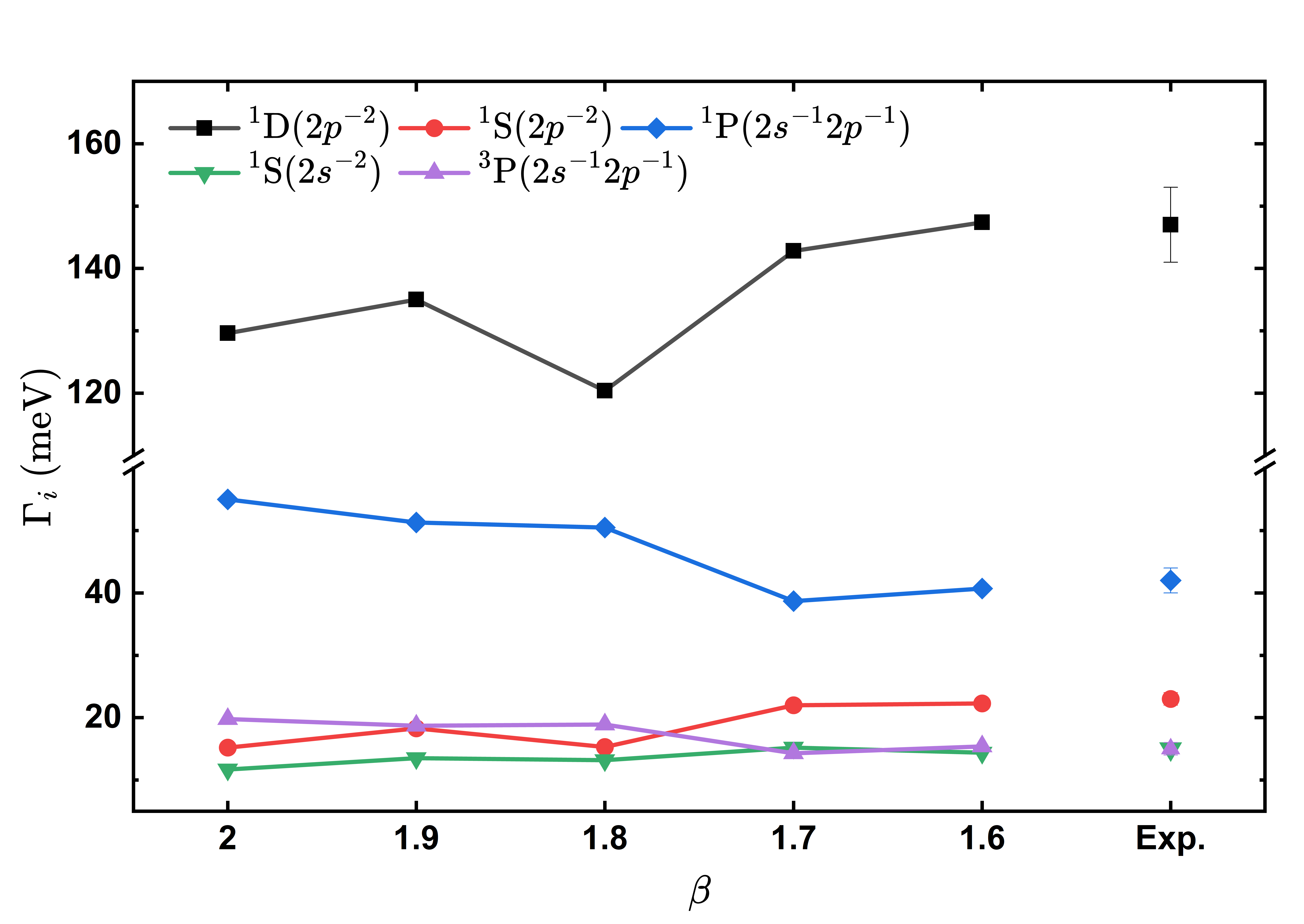}
    \caption{Dependence of the partial Auger decay widths $\Gamma_i$ for the Ne K--LL channels on the even-tempered scaling factor $\beta$ of the continuum SGTO basis. The basis sets are defined in Table~\ref{tab:Ne_beta}}
    \label{fig:Ne_beta}
\end{figure}

\begin{table}
\scriptsize
\caption{
Effect of the scaling factor $\beta$ in the continuum SGTO basis $[N_\beta(spd);\beta,13000]$ on the partial and total Auger decay widths (meV) for Ne K--LL transitions, calculated using the LS-CW-IC model. For each $\beta$, $N_\beta$ is adjusted to cover approximately the same range of Gaussian exponents. The cc-pVQZ basis is used for all bound-state calculations.
}
\label{tab:Ne_beta}
\begin{tabular}{lrrrrrr} 
\hline
\multirow{2}{*}{Channel} & \multicolumn{5}{c}{$\beta\;(N_\beta)$} &  \multirow{2}{*}{Exp.\cite{Albiez1990Z, Muller2017}} \\
\cline{2-6}
& 2.0 (20) 
& 1.9 (21)  
& 1.8 (24) 
& 1.7 (25)
& 1.6 (29)
\\
\hline
$^1$D$(2p^{-2})$
& 129.6 & 135.0 & 120.4 & 142.8 & 147.4 
& 147$\pm$6
\\
$^1$S$(2p^{-2})$
& 15.2 & 18.3 & 15.3 & 22.0 & 22.3 
& 23$\pm$1
\\
$^1$P$(2s^{-1}2p^{-1})$ 
& 55.0 & 51.3 & 50.5 & 38.7 & 40.7
& 42$\pm$2
\\
$^1$S$(2s^{-2})$
& 11.7 & 13.5 & 13.2 & 15.2 & 14.4 
& 15$\pm$1
\\
$^3$P$(2s^{-1}2p^{-1})$
& 19.8 & 18.7 & 18.9 & 14.3 & 15.4 
& 15$\pm$1
\\
\hline
$\sum \Gamma_i$    
& 231.3 & 236.8 & 218.3 & 233.0 & 240.2 
& 242 $\pm$ 10\\  
\hline                  
\end{tabular} 
\end{table}

We next examine the dependence of the calculated widths on the scaling parameter $\beta$ defining the even-tempered continuum basis. The results are summarized in Table~\ref{tab:Ne_beta} and Fig.~\ref{fig:Ne_beta}. In contrast to the weak dependence on the bound-domain basis, the partial widths show a considerably stronger sensitivity to $\beta$. In particular, the more widely spaced sequences with $\beta=1.8$--$2.0$ lead to significant deviations for several channels. With decreasing $\beta$ to 1.7--1.6, the results become more stable and generally agree with the experimental values within their uncertainties.

The poorer performance for the larger scaling factors can be attributed to an insufficient density of Gaussian exponents for representing the Green's-function operator at the high continuum energies of approximately 800~eV encountered in Ne K--LL Auger decay. As discussed in Ref.~\citenum{Mahato2026}, the Green's-function matrix elements exhibits a pronounced dependence on the Gaussian exponent, which cannot be represented quantitatively when the exponent grid is too sparse. This also suggests that a uniformly even-tempered sequence is not necessarily the most efficient construction of the continuum basis. For example, calculations with $\beta=2.0$ can be substantially improved by adding a few functions in the exponent range where the Green's-function matrix elements exhibit their largest amplitudes and variations, approximately $\alpha=1$--$30$ for the electron energy $\sim800\;$eV (see Table~S2 and Fig.~S1 of the Supplementary Information). A more flexible construction could therefore employ a denser exponent distribution in this region and a coarser spacing elsewhere.


Finally, we compare the present results with previous theoretical calculations. A representative selection of different approaches is collected in Table~\ref{tab:Ne_KLL_compare_lit}. Where several variants were reported in the same study, we selected the most accurate or highest-level calculation (comparison of corresponding branching ratios is provided in Table~S11 and Fig.~S2 of the SI). Even for the relatively simple case of Ne K--LL Auger decay, the reported partial widths exhibit considerable variation, illustrating their sensitivity to the treatment of both the correlated bound states and the continuum electron. Among the previous calculations, the results of \citeauthor{Colle1990}~\cite{Colle1990} are closest to the present values. This agreement is expected, since both approaches explicitly describe the continuum within a coupled-channel scattering framework, although \citeauthor{Colle1990} use optimized Hermite--Gaussian continuum functions and a different normalization procedure.
\begin{table*}[t]
\centering
\scriptsize
\setlength{\tabcolsep}{3.2pt}
\renewcommand{\arraystretch}{1.05}

\caption{
Comparison of partial Auger decay widths $\Gamma_i$ (in meV)
for the Ne K--LL channels obtained from different
theoretical calculations.
}
\label{tab:Ne_KLL_compare_lit}

\resizebox{\textwidth}{!}{%
\begin{tabular}{lrrrrrrrrrr}
\hline
\multirow{2}{*}{Channel} & \multicolumn{9}{c}{Theory} &  \multirow{2}{*}{Exp.\cite{Albiez1990Z, Muller2017}} \\
\cline{2-10}
& \phantom{aaaa}Kelly\textsuperscript{\emph{a}}
& \phantom{aaaa}Howat\textsuperscript{\emph{b}}
& \phantom{aa}Nicolaides\textsuperscript{\emph{c}}
& \phantom{aaaaa}Colle\textsuperscript{\emph{d}}
& \phantom{aa}Yarzhemsky\textsuperscript{\emph{e}}
& \phantom{aa}Koloren\v{c}\textsuperscript{\emph{f}}
& \phantom{aa}Skomorowski\textsuperscript{\emph{g}}
& \phantom{aaaa}Matz\textsuperscript{\emph{h}}
& \phantom{aa}Present\textsuperscript{\emph{i}}
\\
\hline
$^1$D$(2p^{-2})$                
& 134.3 & 149.7 & 153.7 & 144.6 & 125.4 & 126.8 & 105.5 & 149.4 & 147.4 & $147\pm6$ \\

$^1$S$(2p^{-2})$                
& 21.0 & 22.6 & 10.9 & 22.0 & 17.7 & 23.8 & 12.2 & 14.0 & 22.3 & $23\pm1$ \\

$^1$P$(2s^{-1}2p^{-1})$         
& 37.2 & 40.8 & 44.9 & 43.4 & 45.1 & 48.9 & 41.5 & 55.4 & 40.7 & $42\pm2$ \\

$^1$S$(2s^{-2})$                
& 13.3 & 12.3 & 11.7 & 13.1 & 17.1 & 24.8 & 28.0 & 18.6 & 14.4 & $15\pm1$ \\

$^3$P$(2s^{-1}2p^{-1})$         
& 13.4 & 19.0 & 8.7 & 19.6 & 13.6 & 26.6 & 19.2 & 13.2 & 15.4 & $15\pm1$ \\
\hline
\end{tabular}%
}
\textsuperscript{\emph{a}}Real-coordinate diagrammatic perturbation theory~\cite{kelly1975}.\\
\textsuperscript{\emph{b}}SCF coupled-channel with transition state basis set~\cite{Howat1978}.\\
\textsuperscript{\emph{c}}Complex-coordinate MC-SCF with mixing of asymptotic pairs (MAP)~\cite{Nicolaides1985}.\\
\textsuperscript{\emph{d}}MC-SCF coupled-channel with optimized Hermite-Gaussian functions for the continuum\cite{Colle1990}.\\
\textsuperscript{\emph{e}}Many-body perturbation theory with frozen approximation and ground state correlation~\cite{Yarzhemsky2002}.\\
\textsuperscript{\emph{f}}Fano-ADC(2)x with uncontracted cc-pCV6Z + 5s5p6d KBJ basis set~\cite{Kolorenc2011}.\\
\textsuperscript{\emph{g}}EOM-CCSD with Coulomb-wave continuum model and $Z_{\textrm{eff}}=6.2$~\cite{Skomorowski2021_2}.\\
\textsuperscript{\emph{h}}Complex-scaling CS-$\Delta$CCSD/aug-cc-pCV5Z with decomposition of the density-matrix CCSD energy formula~\cite{Matz2022}.\\
\textsuperscript{\emph{i}}Present LS-CW-IC results.
\end{table*}

It is also instructive to compare our results with approaches that avoid an explicit  representation of the continuum orbital. The complex-coordinate calculations of  \citeauthor{Nicolaides1985}~\cite{Nicolaides1985}, complex-scaled CCSD calculations of \citeauthor{Matz2022}~\cite{Matz2022}, and Fano-ADC(2)x calculations of \citeauthor{Kolorenc2011}~\cite{Kolorenc2011} provide reasonable results overall, but show larger deviations for several individual channels. This is particularly apparent for the relative widths of the two $^1$S channels. These results illustrate that an accurate total resonance width does not necessarily imply an equally accurate channel-resolved decomposition, for which the explicit treatment of the continuum and interchannel coupling can be important. The present LS-CW-IC results also substantially improve upon the previous EOM-CCSD calculations employing a continuum model with  the Coulomb wave\cite{Skomorowski2021_2}. While retaining the same EOM-CCSD treatment of the bound states, solving the LS equation and explicitly including interchannel coupling considerably improves the individual partial widths.

Overall, Table~\ref{tab:Ne_KLL_compare_lit} demonstrates that the LS solution with SGTO representation introduced here provides an effective way of combining an explicit scattering description of the continuum orbital with correlated many-electron wave functions, yielding consistently accurate partial Auger widths without optimized basis set functions or any empirical parametrization.

\subsection{K-shell decay in \ce{H2O}}
We next examine the performance of the present method for molecules, starting with K-shell Auger decay in water molecule, whose Auger spectrum has been extensively studied both experimentally and theoretically and is frequently used as a benchmark for theoretical calculations\cite{Moddeman1971, Siegbahn1975, Agren1980, Carravetta1987,  Cesar1989, Ohrwall2025, Takahashi2006, Inhester2012, Inhester2014, Matz2022, Skomorowski2025}. Although \ce{H2O} is isoelectronic with Ne and its K-shell Auger decay largely resembles the corresponding Ne K--LL process, its lower symmetry results in a much larger number of symmetry-distinct final ionic states. In addition, nuclear motion and electronic correlation make the molecular Auger spectrum considerably more challenging to describe theoretically.



Extending the present method from an atom to molecules introduces two additional aspects that require separate consideration: the choice of the continuum basis set and the definition of the asymptotic Coulomb potential in the LS-CW model. For the continuum basis, we adopt the following strategy. On the atom containing the core hole (oxygen in the case of \ce{H2O}), we place a full even-tempered SGTO basis, constructed in the same manner as for Ne. This basis is supplemented with uncontracted Gaussian functions on the hydrogen atoms, taken from the corresponding bound-state basis set. This choice reflects the expectation that, for calculating Auger decay widths, an accurate representation of the continuum orbital near the core-ionized atom is particularly important. 
In the following we use the notation O[...] + H[...] to specify the continuum basis functions placed on the respective atomic centers.

The second aspect concerns the choice of the Coulomb center defining the asymptotic potential in the LS-CW formulation. Unlike in an isolated atom, where the Coulomb center is naturally located at the nucleus, its position in a molecule without a center of symmetry is not uniquely defined. 
Here we consider two physically motivated choices: placing the Coulomb center at the oxygen nucleus, consistent with the construction of the continuum basis, or at the origin of the coordinate system, which coincides with the center of nuclear charge in the molecule. In principle, the exact scattering solution should be independent of this choice, since different partitions of the Hamiltonian into reference and interaction potentials are formally equivalent. However, in the present implementation, the Coulomb Green's function is represented through a Dyson equation projected onto a finite Gaussian basis. Therefore, this approximation may introduce some dependence of the results on the choice of the Coulomb center.

\begin{table*}[t]
\centering
\scriptsize
\setlength{\tabcolsep}{3.2pt}
\renewcommand{\arraystretch}{1.05}
\caption{
Auger electron energies (in eV) and partial decay widths  $\Gamma_i$ (in meV) for the main K--LL Auger decay channels of \ce{H2O}, calculated using the cc-pVQZ / O$[29(spd);1.6,13000]$ $+$ H[un-cc-pVQZ] basis sets. In the LS-CW calculations, the Coulomb wave center is placed either at the oxygen atom (O) or at the origin of the coordinate system (0). In the LS-CW-IC and LS-PW-IC models, interchannel coupling between singlet and triplet channels is neglected.}
\label{tab:H2O_ccpvqz_29}
\resizebox{\textwidth}{!}{%
\begin{tabular}{lcrrrrrrrr}
\hline
Channel & Auger Energy & PW & Proj-PW & LS-PW
& LS-PW-IC & LS-CW(O) & LS-CW-IC(O)
& LS-CW(0) & LS-CW-IC(0) \\
\hline
$^1$A$_1$(1b$_1^{-2}$)
& 498.0 & 17.8 & 17.8 & 17.2 & 18.9 & 17.5 & 19.2 & 17.7 & 19.4 \\

$^1$B$_1$(3a$_1^{-1}$1b$_1^{-1}$)
& 496.5 & 18.5 & 18.5 & 16.2 & 17.3 & 17.3 & 18.4 & 16.9 & 17.9 \\

$^1$A$_1$(3a$_1^{-2}$)
& 493.4 & 11.1 & 11.2 & 11.7 & 13.2 & 11.1 & 12.5 & 12.0 & 13.5 \\

$^1$A$_2$(1b$_1^{-1}$1b$_2^{-1}$)
& 493.2 & 16.6 & 16.6 & 13.7 & 14.6 & 14.2 & 15.1 & 14.5 & 15.4 \\

$^1$B$_2$(3a$_1^{-1}$1b$_2^{-1}$)
& 491.0 & 13.5 & 13.6 & 11.9 & 12.7 & 12.7 & 13.6 & 12.3 & 13.1 \\

$^1$A$_1$(1b$_2^{-2}$)
& 486.3 &  6.9 &  6.8 &  9.3 & 10.8 &  8.0 &  9.3 &  9.5 & 11.0 \\

$^1$B$_1$(2a$_1^{-1}$1b$_1^{-1}$)
& 473.7 &  1.7 &  1.7 & 11.4 &  9.7 &  9.3 &  7.7 & 11.1 &  9.4 \\

$^1$A$_1$(2a$_1^{-1}$3a$_1^{-1}$)
& 472.4 &  3.7 &  3.7 & 13.4 & 11.7 & 12.0 & 10.5 & 12.9 & 11.2 \\

$^1$B$_2$(2a$_1^{-1}$1b$_2^{-1}$)
& 467.5 &  1.4 &  1.4 &  7.6 &  6.6 &  6.3 &  5.4 &  7.6 &  6.6 \\

$^1$A$_1$(2a$_1^{-2}$)
& 453.9 &  9.6 &  9.6 & 10.7 &  8.4 &  6.0 &  4.5 & 10.0 &  7.8 \\

$^3$B$_1$(3a$_1^{-1}$1b$_1^{-1}$)
& 498.9 &  3.3 &  3.3 &  0.5 &  0.4 &  0.5 &  0.4 &  0.4 &  0.4 \\

$^3$A$_2$(1b$_1^{-1}$1b$_2^{-1}$)
& 494.9 &  0.0 &  0.0 &  0.0 &  0.0 &  0.0 &  0.0 &  0.0 &  0.0 \\

$^3$B$_2$(3a$_1^{-1}$1b$_2^{-1}$)
& 493.1 &  2.5 &  2.5 &  0.4 &  0.3 &  0.3 &  0.3 &  0.3 &  0.3 \\

$^3$B$_1$(2a$_1^{-1}$1b$_1^{-1}$)
& 480.4 & 24.5 & 24.5 &  3.9 &  3.8 &  3.9 &  3.7 &  3.7 &  3.6 \\

$^3$A$_1$(2a$_1^{-1}$3a$_1^{-1}$)
& 478.6 & 23.2 & 23.1 &  3.7 &  3.6 &  3.5 &  3.4 &  3.5 &  3.4 \\

$^3$B$_2$(2a$_1^{-1}$1b$_2^{-1}$)
& 474.8 & 18.0 & 18.0 &  2.9 &  2.8 &  2.9 &  2.8 &  2.8 &  2.7 \\

\hline
$\sum \Gamma_i$
& & 172.1 & 172.3 & 134.6 & 134.9
& 125.3 & 126.8 & 135.3 & 136.0 \\
\hline
\end{tabular}%
}
\end{table*}
Table~\ref{tab:H2O_ccpvqz_29} presents the calculated partial widths for the main KLL Auger decay channels of \ce{H2O}, obtained using the different continuum orbital models as discussed above for Ne. First, we note that the projected plane-wave (Proj-PW) results are virtually identical to those obtained with exact plane waves (PW) for all channels, confirming that the chosen continuum basis provides an accurate representation of the continuum orbital in the region relevant to the Auger transition. Including the interaction potential through the Lippmann--Schwinger equation has the most pronounced effect on the triplet channels, whose partial widths are substantially reduced compared with the PW results. In contrast, the widths of several singlet channels increase significantly, particularly those leading to final dicationic states with one inner-valence (2a$_1^{-1}$) hole and one outer-valence (1b$_1^{-1}$, 1b$_2^{-1}$, or 3a$_1^{-1}$) hole.
As in Ne, the differences between the LS-PW and LS-CW formulations are generally modest, as is the effect of interchannel coupling. A notable exception is the $^1$A$_1(2$a$_1^{-2})$ channel, whose partial width is particularly sensitive to both interchannel coupling and the choice of the Coulomb center. For this channel, the width decreases from 10.7~meV in the LS-PW model to 4.5~meV in the LS-CW-IC(O) model. 
As in Ne, the inclusion of the orthogonalizing pseudopotential (OPP) has a negligible effect on the calculated partial widths (see Table~S12 of the SI). Similarly, the contribution of continuum basis functions centered on the hydrogen atoms is very small. As shown by comparing the results in Table~\ref{tab:H2O_ccpvqz_29} with those in Table~S13 of the SI, the inclusion of hydrogen-centered continuum functions leaves most partial widths unchanged, affecting only a few channels by no more than 0.3~meV. This suggests that, for molecular Auger decay, it may be sufficient to place continuum basis functions only on the atom (or atoms) containing the core hole, even in larger molecules. This observation is also consistent with the success of the one-center approximation in calculations of Auger decay widths.

We now compare the present results with previous theoretical calculations of Auger decay in \ce{H2O}. Although this system has been extensively studied, we focus primarily on approaches employing the same underlying electronic-structure method (EOM-CCSD or CCSD) to better isolate the effect of the continuum representation. Table~\ref{tab:H2O_compare_lit} compares the calculated partial widths, while the corresponding relative widths are reported in Table~S14 of the SI. As a reference, we also include the results of \citeauthor{Inhester2012}~\cite{Inhester2012, Inhester2014}, who combined MRCI calculations of the bound states with a numerical solution of the Schr\"odinger equation for the continuum electron. These calculations can be regarded the most comprehensive theoretical treatments of channel-resolved Auger decay in \ce{H2O} reported to date. For consistency with the center of partial-wave expansion employed in Ref.~\citenum{Inhester2012}, the Coulomb center in our calculations is placed at the oxygen atom.

Inspection of Table~\ref{tab:H2O_compare_lit} reveals excellent agreement between the present LS-CW-IC(O) results and those of \citeauthor{Inhester2012} for nearly all channels. This agreement is particularly noteworthy given the very different representations of the continuum electron. In the present approach, the continuum orbital is obtained by solving the Lippmann--Schwinger equation in a finite SGTO basis, whereas \citeauthor{Inhester2012} solved the Schr\"odinger equation numerically, imposing the appropriate asymptotic boundary conditions. Despite these differences, the two approaches yield remarkably similar partial widths, providing strong support for the accuracy of the present continuum representation.

\begin{table}[t!]
\footnotesize
\setlength{\tabcolsep}{4.0pt}
\renewcommand{\arraystretch}{1.05}
\caption{\label{tab:H2O_compare_lit}
Comparison of partial Auger decay widths $\Gamma_i$ (in meV) for the main channels in the  K--LL Auger decay of core-ionized water obtained using different theoretical approaches.}

\begin{tabular}{lrrrrr}
\hline
\multirow{2}{*}{Channel}
& \multicolumn{4}{c}{EOM-CCSD}
& \multirow{2}{*} {MRCI\textsuperscript{\emph{e}}} \\
\cline{2-5}
&  CW\textsuperscript{\emph{a}}
& CBF\textsuperscript{\emph{b}}
& OCA\textsuperscript{\emph{c}}
& Present\textsuperscript{\emph{d}}
\\
\hline

$^1$A$_1$(1b$_1^{-2}$)
& 13.3 & 18.0 & 27.3 & 19.2 & 19.0 \\

$^1$B$_1$(3a$_1^{-1}$1b$_1^{-1}$)
& 12.7 & 19.6 & 25.7 & 18.4 & 18.0 \\

$^1$A$_1$(3a$_1^{-2}$)
&  8.9 & 12.2 & 17.2 & 12.5 & 13.1 \\

$^1$A$_2$(1b$_1^{-1}$1b$_2^{-1}$)
& 10.7 & 15.7 & 18.4 & 15.1 & 15.2 \\

$^1$B$_2$(3a$_1^{-1}$1b$_2^{-1}$)
&  9.5 & 13.4 & 15.8 & 13.6 & 13.2 \\

$^1$A$_1$(1b$_2^{-2}$)
&  7.1 &  8.7 &  9.2 &  9.3 &  9.8 \\

$^1$B$_1$(2a$_1^{-1}$1b$_1^{-1}$)
&  9.5 &  9.6 & 15.4 &  7.7 & 10.0 \\

$^1$A$_1$(2a$_1^{-1}$3a$_1^{-1}$)
& 13.6 & 12.7 & 18.2 & 10.5 & 11.0 \\

$^1$B$_2$(2a$_1^{-1}$1b$_2^{-1}$)
&  6.3 &  6.8 &  7.9 &  5.4 &  6.6 \\

$^1$A$_1$(2a$_1^{-2}$)
& 15.3 & 21.6 &  9.4 &  4.5 &  4.1 \\

$^3$B$_1$(3a$_1^{-1}$1b$_1^{-1}$)
&  0.5 &  0.2 &  1.0 &  0.4 &  0.4 \\

$^3$A$_2$(1b$_1^{-1}$1b$_2^{-1}$)
&  0.0 &  0.0 &  0.0 &  0.0 &  0.0 \\

$^3$B$_2$(3a$_1^{-1}$1b$_2^{-1}$)
&  0.4 &  0.2 &  0.7 &  0.3 &  0.3 \\

$^3$B$_1$(2a$_1^{-1}$1b$_1^{-1}$)
&  4.1 &  2.8 &  3.4 &  3.7 &  3.0 \\

$^3$A$_1$(2a$_1^{-1}$3a$_1^{-1}$)
&  3.8 &  2.5 &  3.3 &  3.4 &  2.6 \\

$^3$B$_2$(2a$_1^{-1}$1b$_2^{-1}$)
&  2.9 &  2.2 &  2.1 &  2.8 &  1.6 \\

\hline
\end{tabular}

\textsuperscript{\emph{a}} Coulomb-wave continuum model with $Z_{\mathrm{eff}}=4.9$~\cite{Skomorowski2021_2}.
\\
\textsuperscript{\emph{b}} Complex basis function $\Delta$CCSD approach with uncontracted 6-311+G(3df) basis set~\cite{Matz2022}.
\\
\textsuperscript{\emph{c}} One-center approximation with STO-3G minimal basis set~\cite{Skomorowski2025}.
\\
\textsuperscript{\emph{d}} LS-CW-IC(O) results with the cc-pVQZ / O$[29(spd);1.6,13000]$ + H[un-cc-pVQZ] basis sets
\\
\textsuperscript{\emph{e}} MRCI with numerical continuum orbital~\cite{Inhester2012}.
\end{table}


The comparison with other EOM-CCSD-based approaches reveals more pronounced differences. The Coulomb-wave model (CW) with a fixed effective charge~\cite{Skomorowski2021_2} provides only semiquantitative agreement with both the MRCI reference and the present LS-CW-IC(O) results, with noticeable deviations in the relative widths of several singlet channels. The one-center approximation (OCA)~\cite{Skomorowski2025}, on the other hand, reproduces the relative distribution of partial widths reasonably well, but systematically overestimates their absolute values compared with both sets of results. The complex-basis-function (CBF) approach generally agrees well with both the MRCI and present results, although it shows a large discrepancy for the $^1$A$_1(2$a$_1^{-2})$ channel. An additional limitation of the CBF approach is the pronounced basis-set dependence reported in Ref.~\citenum{Matz2022}.

\begin{figure}[h!]
\includegraphics[width=0.9\linewidth]{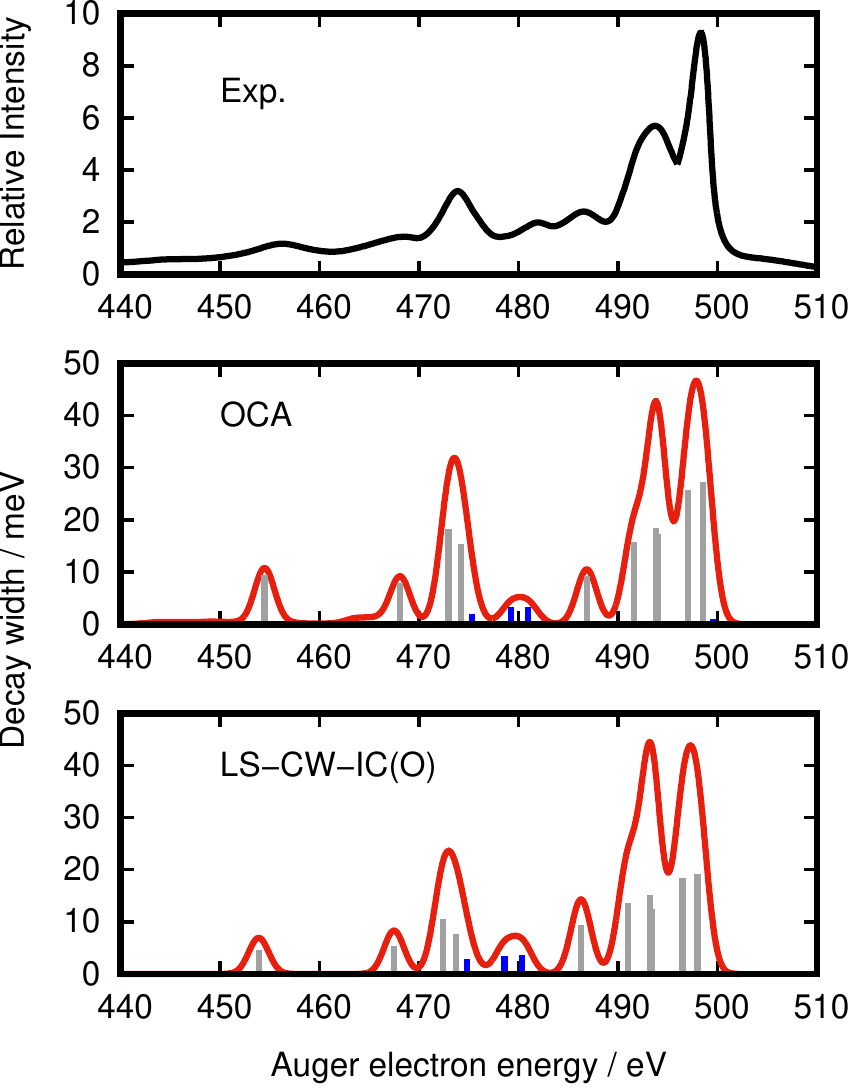}
\caption{
 Experimental~\cite{Moddeman1971} and calculated normal Auger spectra of singly ionized \ce{H2O}. The theoretical spectra were obtained by applying Gaussian broadening with a full width at half maximum (FWHM) of 1.0~eV to the stick spectra. Gray and blue sticks represent singlet and triplet decay channels, respectively. The OCA spectrum is taken from Ref.~\citenum{Skomorowski2025}, while the present results were obtained using the LS-CW-IC(O) model.
 }
\label{fig:KLL-h2o}
\end{figure}
Finally, we compare the calculated Auger spectra with experiment. Figure~\ref{fig:KLL-h2o} shows the spectra obtained using the present LS-CW-IC(O) model and the one-center approximation (OCA), together with the experimental spectrum~\cite{Moddeman1971}. Both theoretical spectra were generated by applying a uniform Gaussian broadening of 1~eV (FWHM) to the discrete stick spectra. This simplified treatment mimics the effects of the nuclear motion, which contributes substantially to the broadening observed in the experimental spectrum.
Despite neglecting explicit nuclear dynamics, the present LS-CW-IC(O) approach reproduces the main features of the experimental spectrum, including the dominant intensity in the high-energy region and the weaker structures at lower energies. A more accurate description of the relative intensities of the broad peaks around 495--500~eV is expected to require explicit treatment of vibrational Franck--Condon effects, as demonstrated in Ref.~\citenum{Inhester2012}.

Interestingly, the LS-CW-IC(O) and OCA spectra are remarkably similar, despite the substantial differences in their absolute partial widths (Table~\ref{tab:H2O_compare_lit}). This reflects the much closer agreement between their relative widths, which primarily determine the spectral shape. Nevertheless, the LS-CW-IC(O) spectrum provides a slightly better description of the relative intensities in the intermediate- and low-energy regions, particularly for transitions involving an inner-valence 2a$_1^{-1}$ hole in the final dicationic state. Overall, the differences between the two theoretical spectra are minor, suggesting that further improvements in the agreement with experiment will primarily require a more accurate treatment of nuclear dynamics rather than further refinement of the continuum representation.

\subsection{K-shell decay in \ce{CO2}}
We next consider the Auger decay of core-ionized \ce{CO2}, another widely studied benchmark system in core-level spectroscopy\cite{Moddeman1971, Kelber1981, Agren1981, Hiltunen1999, Feyer2005, Puttner2008, Eland2016, Skomorowski2025}. Depending on whether the initial core vacancy is created at the carbon or oxygen K edge, \ce{CO2} exhibits two distinct KLL Auger spectra, both of which are examined here. 
As a larger molecular system than \ce{H2O}, \ce{CO2} provides a useful test of the applicability of the present approach to more complex molecules. Our analysis focuses on two aspects: the dependence of the calculated Auger widths on the continuum basis set and their comparison with results obtained using the one-center approximation (OCA).

\begin{table*}[ht!]
\footnotesize
\setlength{\tabcolsep}{4.0pt}
\renewcommand{\arraystretch}{1.05}
\caption{\label{tab:CO2_C_O_comparison_no_f}
Double ionization potential (EOM-DIP) energies (in eV) and partial Auger decay widths ($\Gamma_i$, in meV) for the main channels in the Auger decay of K-shell-ionized CO$_2$ at the carbon and oxygen  K-edges. The cc-pVTZ basis set is used for the bound states. In the LS calculations, the SGTO continuum basis is C$[27(spd);1.6,5078.1]$ + O[un-cc-pVTZ] for the C edge, and O$[27(spd);1.6,5078.1]$ + C[un-cc-pVTZ] for the O edge. One-center approximation (OCA) results are included for comparison.}

\begin{tabular}{lccccccc}
\hline
\multirow{2}{*}{Channel}
& \multirow{2}{*}{EOM-DIP}
& \multicolumn{3}{c}{C edge}
& \multicolumn{3}{c}{O edge} \\
\cmidrule(r){3-5}\cmidrule(l){6-8}
& & \phantom{ }PW\phantom{ } & LS-CW-IC & OCA\cite{Skomorowski2025}
& \phantom{ }PW\phantom{ } & LS-CW-IC & OCA\cite{Skomorowski2025} \\
\hline

$^1\Delta_{\mathrm{g}}$(1$\pi_{\mathrm{g}}^{-2}$)
& 38.9 & 1.1 & 0.7 & 1.1 & 9.0 & 7.6 & 12.9 \\

$^1\Sigma_{\mathrm{g}}^{+}$(1$\pi_{\mathrm{g}}^{-2}$)
& 39.4 & 0.3 & 0.3 & 0.4 & 1.8 & 3.9 & 4.3 \\

$^1\Pi_{\mathrm{u}}$(3$\sigma_{\mathrm{u}}^{-1}$1$\pi_{\mathrm{g}}^{-1}$)
& 42.7 & 0.3 & 0.8 & 0.8 & 4.1 & 3.7 & 5.6 \\

$^1\Pi_{\mathrm{g}}$(4$\sigma_{\mathrm{g}}^{-1}$1$\pi_{\mathrm{g}}^{-1}$)
& 43.3 & 2.5 & 1.7 & 1.6 & 0.6 & 0.8 & 1.7 \\

$^1\Delta_{\mathrm{u}}$(1$\pi_{\mathrm{u}}^{-1}$1$\pi_{\mathrm{g}}^{-1}$)
& 45.6 & 0.3 & 0.0 & 0.0 & 15.2 & 12.6 & 16.9 \\

$^1\Sigma_{\mathrm{g}}^{+}$(3$\sigma_{\mathrm{u}}^{-2}$)
& 45.9 & 1.1 & 2.4 & 1.7 & 0.3 & 0.6 & 0.4 \\

$^1\Sigma_{\mathrm{u}}$(1$\pi_{\mathrm{u}}^{-1}$1$\pi_{\mathrm{g}}^{-1}$)
& 46.7 & 0.1 & 0.1 & 0.1 & 3.7 & 8.1 & 7.4 \\

$^1\Delta_{\mathrm{g}}$(1$\pi_{\mathrm{u}}^{-2}$)
& 48.0 & 9.6 & 6.7 & 7.5 & 9.0 & 7.4 & 9.3 \\

$^1\Sigma_{\mathrm{g}}^{+}$(1$\pi_{\mathrm{u}}^{-2}$)
& 48.8 & 2.2 & 3.1 & 3.1 & 3.0 & 5.5 & 5.2 \\

$^1\Pi_{\mathrm{g}}$(3$\sigma_{\mathrm{u}}^{-1}$1$\pi_{\mathrm{u}}^{-1}$)
& 49.7 & 5.2 & 3.3 & 2.9 & 12.3 & 11.9 & 16.1 \\

$^1\Pi_{\mathrm{u}}$(4$\sigma_{\mathrm{g}}^{-1}$1$\pi_{\mathrm{u}}^{-1}$)
& 50.4 & 0.3 & 3.1 & 2.4 & 10.3 & 10.0 & 13.2 \\

$^1\Sigma_{\mathrm{u}}$(4$\sigma_{\mathrm{g}}^{-1}$3$\sigma_{\mathrm{u}}^{-1}$)
& 54.7 & 0.1 & 2.0 & 1.1 & 3.5 & 5.3 & 5.7 \\

$^1\Sigma_{\mathrm{g}}^{+}$(4$\sigma_{\mathrm{g}}^{-2}$)
& 55.1 & 2.9 & 0.9 & 0.4 & 3.1 & 4.8 & 5.1 \\

$^1\Sigma_{\mathrm{g}}^{+}$(2$\sigma_{\mathrm{u}}^{-1}$3$\sigma_{\mathrm{u}}^{-1}$)
& 63.9 & 2.3 & 5.3 & 2.0 & 0.1 & 0.2 & 0.2 \\

$^1\Pi_{\mathrm{g}}$(2$\sigma_{\mathrm{u}}^{-1}$1$\pi_{\mathrm{u}}^{-1}$)
& 71.3 & 4.1 & 2.4 & 1.4 & 1.5 & 5.3 & 6.5 \\

$^1\Pi_{\mathrm{u}}$(3$\sigma_{\mathrm{g}}^{-1}$1$\pi_{\mathrm{u}}^{-1}$)
& 71.8 & 0.7 & 3.4 & 1.0 & 1.5 & 5.9 & 7.3 \\

$^1\Sigma_{\mathrm{u}}$(3$\sigma_{\mathrm{g}}^{-1}$3$\sigma_{\mathrm{u}}^{-1}$)
& 73.5 & 0.2 & 1.7 & 0.7 & 3.2 & 9.7 & 8.8 \\

$^1\Sigma_{\mathrm{g}}^{+}$(3$\sigma_{\mathrm{g}}^{-1}$4$\sigma_{\mathrm{g}}^{-1}$)
& 73.9 & 1.9 & 1.3 & 0.3 & 1.6 & 4.5 & 4.1 \\

$^3\Pi_{\mathrm{u}}$(3$\sigma_{\mathrm{u}}^{-1}$1$\pi_{\mathrm{g}}^{-1}$)
& 42.0 & 1.5 & 0.2 & 0.1 & 1.4 & 0.2 & 0.6 \\

$^3\Pi_{\mathrm{g}}$(4$\sigma_{\mathrm{g}}^{-1}$1$\pi_{\mathrm{g}}^{-1}$)
& 43.0 & 0.1 & 0.0 & 0.0 & 2.1 & 0.3 & 0.5 \\

$^3\Pi_{\mathrm{g}}$(3$\sigma_{\mathrm{u}}^{-1}$1$\pi_{\mathrm{u}}^{-1}$)
& 47.5 & 0.2 & 0.0 & 0.0 & 4.7 & 0.6 & 1.3 \\

$^3\Pi_{\mathrm{u}}$(4$\sigma_{\mathrm{g}}^{-1}$1$\pi_{\mathrm{u}}^{-1}$)
& 48.3 & 8.7 & 1.6 & 0.8 & 4.3 & 0.5 & 1.0 \\

\hline
$\sum \Gamma_i$
& & 45.7 & 41.1 & 29.2 & 96.1 & 109.5 & 134.0 \\
\hline
\end{tabular}
\end{table*}

Table~\ref{tab:CO2_C_O_comparison_no_f} presents selected results for the most intense Auger decay channels at both carbon and oxygen K edges, with the O-edge widths averaged over the two initial core-ionized states. A more detailed comparison of the different continuum models for both edges is provided in the Supporting Information (Tables~S15--S18). As in the preceding systems, the largest changes in partial widths with respect to the plane wave model arise from including the scattering potential through the LS-PW formulation, whereas the additional effects of the LS-CW formulation and interchannel coupling are relatively small. The scattering potential has the strongest influence on triplet channels, whose widths are substantially reduced relative to the PW results, while singlet channels are generally less affected. Overall, the influence of the scattering potential is less pronounced than in Ne or \ce{H2O}.

We also compare in Table~\ref{tab:CO2_C_O_comparison_no_f} the present LS-CW-IC results with those obtained using the one-center approximation (OCA), employing the same underlying EOM-CCSD description of the bound states. Although the OCA and LS-CW-IC results exhibit noticeable differences in absolute partial widths, they generally agree on the dominant decay channels for both edges. The OCA widths tend to be smaller than the LS-CW-IC values for the C edge, whereas the opposite trend is observed for the O edge.

\begin{table*}[!t]
\centering
\setlength{\tabcolsep}{4.0pt}
\renewcommand{\arraystretch}{1.05}
\setlength{\tabcolsep}{5pt}
\renewcommand{\arraystretch}{1.12}
\caption{\label{tab:CO2_O_LSPW_basis}
Dependence of the partial Auger decay widths (in meV) on the continuum basis set for the oxygen K edge of CO$_2$, calculated using the LS-PW model. The reference basis is $\mathrm{O}[27(spd);\beta,5078.1]+\mathrm{C}[\text{un-cc-pVTZ}]$. The "$+f$" variant additionally includes $\mathrm{O}[5(f);1.6,7.0]$, and "No C" variant omits the carbon-centered continuum functions\textsuperscript{\emph{a}}. Double ionization potential (EOM-DIP) energies (in eV) are also given}
\begin{tabular}{lcrrrr}
\hline
\multirow{2}{*}{Channel} & \multirow{2}{*}{EOM-DIP} & \multicolumn{3}{c}{$\beta = 1.6$} & \phantom{a}$\beta =2.0$   \\
\cmidrule(r){3-5}
\cmidrule(l){6-6}
& & Reference & $+f$ & No C & No C\\
\hline
$^1\Delta_{\mathrm{g}}$(1$\pi_{\mathrm{g}}^{-2}$) & 38.9 & 7.4 & 7.4 & 7.4 & 6.6 \\
$^1\Sigma_{\mathrm{g}}^{+}$(1$\pi_{\mathrm{g}}^{-2}$) & 39.4 & 2.6 & 2.5 & 2.5 & 2.3 \\
$^1\Pi_{\mathrm{u}}$(3$\sigma_{\mathrm{u}}^{-1}$1$\pi_{\mathrm{g}}^{-1}$) & 42.7 & 3.5 & 3.5 & 3.5 & 3.2 \\
$^1\Pi_{\mathrm{g}}$(4$\sigma_{\mathrm{g}}^{-1}$1$\pi_{\mathrm{g}}^{-1}$) & 43.3 & 0.8 & 0.8 & 0.8 & 0.8 \\
$^1\Delta_{\mathrm{u}}$(1$\pi_{\mathrm{u}}^{-1}$1$\pi_{\mathrm{g}}^{-1}$) & 45.6 & 12.3 & 12.3 & 12.3 & 11.2 \\
$^1\Sigma_{\mathrm{g}}^{+}$(3$\sigma_{\mathrm{u}}^{-2}$) & 45.9 & 0.4 & 0.4 & 0.4 & 0.4 \\
$^1\Sigma_{\mathrm{u}}$(1$\pi_{\mathrm{u}}^{-1}$1$\pi_{\mathrm{g}}^{-1}$) & 46.7 & 5.5 & 5.2 & 5.3 & 5.0 \\
$^1\Delta_{\mathrm{g}}$(1$\pi_{\mathrm{u}}^{-2}$) & 48.0 & 7.2 & 7.3 & 7.2 & 6.7 \\
$^1\Sigma_{\mathrm{g}}^{+}$(1$\pi_{\mathrm{u}}^{-2}$) & 48.8 & 3.9 & 3.7 & 3.8 & 3.6 \\
$^1\Pi_{\mathrm{g}}$(3$\sigma_{\mathrm{u}}^{-1}$1$\pi_{\mathrm{u}}^{-1}$) & 49.7 & 11.3 & 11.3 & 11.3 & 10.5 \\
$^1\Pi_{\mathrm{u}}$(4$\sigma_{\mathrm{g}}^{-1}$1$\pi_{\mathrm{u}}^{-1}$) & 50.4 & 9.5 & 9.4 & 9.4 & 8.8 \\
$^1\Sigma_{\mathrm{u}}$(4$\sigma_{\mathrm{g}}^{-1}$3$\sigma_{\mathrm{u}}^{-1}$) & 54.7 & 4.0 & 4.2 & 4.1 & 3.8 \\
$^1\Sigma_{\mathrm{g}}^{+}$(4$\sigma_{\mathrm{g}}^{-2}$) & 55.1 & 3.6 & 3.7 & 3.7 & 3.4 \\
$^1\Sigma_{\mathrm{g}}^{+}$(2$\sigma_{\mathrm{u}}^{-1}$3$\sigma_{\mathrm{u}}^{-1}$) & 63.9 & 0.2 & 0.2 & 0.2 & 0.2 \\
$^1\Pi_{\mathrm{g}}$(2$\sigma_{\mathrm{u}}^{-1}$1$\pi_{\mathrm{u}}^{-1}$) & 71.3 & 5.8 & 5.8 & 5.8 & 5.2 \\
$^1\Pi_{\mathrm{u}}$(3$\sigma_{\mathrm{g}}^{-1}$1$\pi_{\mathrm{u}}^{-1}$) & 71.8 & 6.4 & 6.4 & 6.4 & 5.7 \\
$^1\Sigma_{\mathrm{u}}$(3$\sigma_{\mathrm{g}}^{-1}$3$\sigma_{\mathrm{u}}^{-1}$) & 73.5 & 6.9 & 7.2 & 7.0 & 6.8 \\
$^1\Sigma_{\mathrm{g}}^{+}$(3$\sigma_{\mathrm{g}}^{-1}$4$\sigma_{\mathrm{g}}^{-1}$) & 73.9 & 3.3 & 3.4 & 3.3 & 3.2 \\
$^3\Pi_{\mathrm{u}}$(3$\sigma_{\mathrm{u}}^{-1}$1$\pi_{\mathrm{g}}^{-1}$) & 42.0 & 0.2 & 0.2 & 0.2 & 0.2 \\
$^3\Pi_{\mathrm{g}}$(4$\sigma_{\mathrm{g}}^{-1}$1$\pi_{\mathrm{g}}^{-1}$) & 43.0 & 0.3 & 0.3 & 0.3 & 0.3 \\
$^3\Pi_{\mathrm{g}}$(3$\sigma_{\mathrm{u}}^{-1}$1$\pi_{\mathrm{u}}^{-1}$) & 47.5 & 0.6 & 0.6 & 0.6 & 0.6 \\
$^3\Pi_{\mathrm{u}}$(4$\sigma_{\mathrm{g}}^{-1}$1$\pi_{\mathrm{u}}^{-1}$) & 48.3 & 0.6 & 0.6 & 0.6 & 0.6 \\
\hline
$\sum \Gamma_i$ & & 96.5 & 96.6 & 96.3 & 88.9 \\
\hline
\end{tabular}

\smallskip
\textsuperscript{\emph{a}} A single $s$ function on C atom is added for technical reasons.
\end{table*}

The convergence of the calculated partial widths with respect to the continuum basis set is examined in Table~\ref{tab:CO2_O_LSPW_basis} for the oxygen K-edge, with the corresponding carbon K-edge results provided in the Supporting Information (Table~S19). The oxygen edge is particularly interesting because the presence of two equivalent oxygen atoms requires continuum basis functions on both centers. Inspection of Table~\ref{tab:CO2_O_LSPW_basis} shows that the inclusion of $f$ functions has a negligible effect, indicating that angular momenta up to $d$, as used for the isolated atoms, are sufficient. Similarly, removing carbon-centered continuum functions produces only minor changes. Increasing the even-tempered exponent ratio $\beta$ from 1.6 to 2.0 has a somewhat larger effect, reducing the total width by approximately 8\%. For the carbon K-edge, the sensitivity to the exponent ratio is considerably smaller, with the total width changing by less than 3\%. Overall, these results suggest that, even for larger molecules, an accurate continuum representation can be achieved using a relatively compact basis localized exclusively on the core-ionized atom or equivalent core-ionization centers.

\begin{figure*}[t!]
\includegraphics[width=0.45\linewidth]{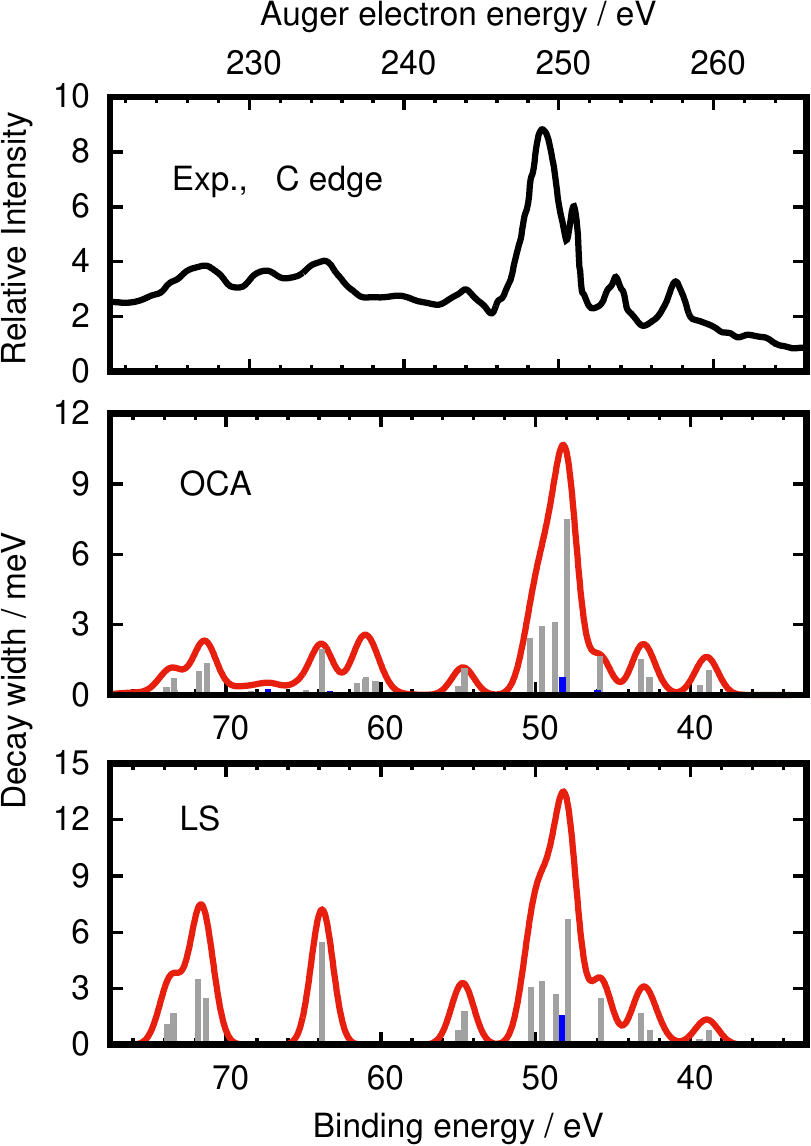}
\includegraphics[width=0.435\linewidth]{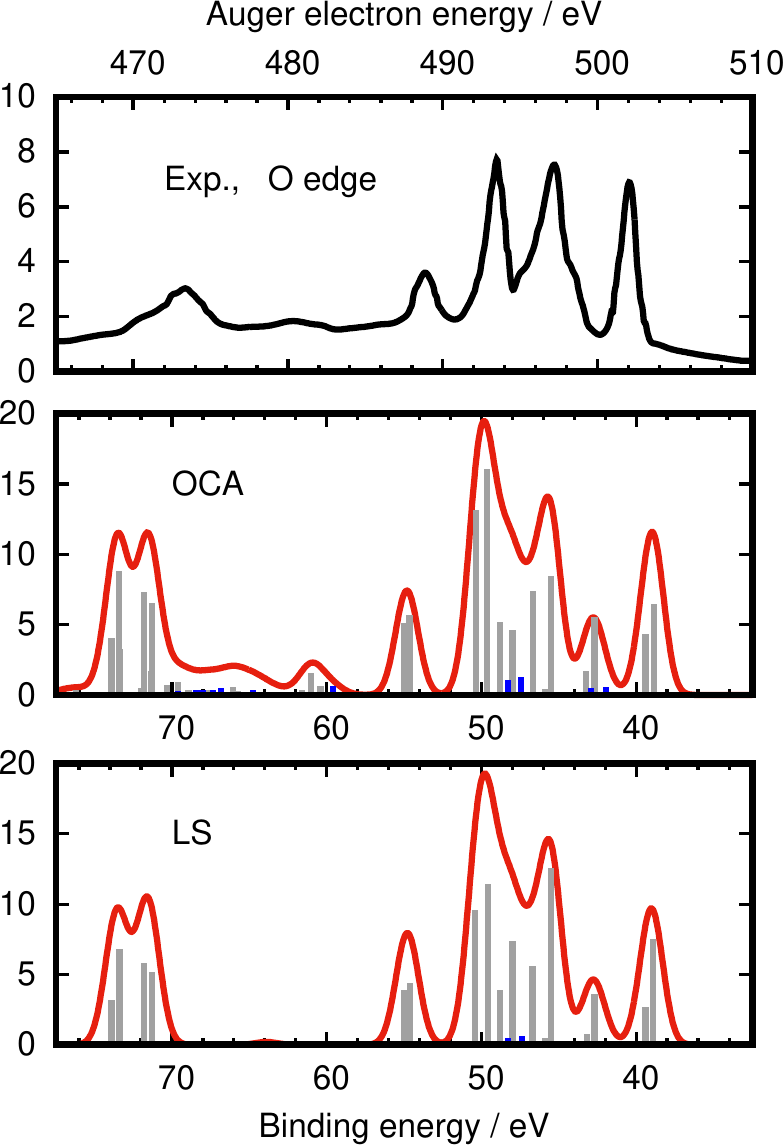}
\caption{
 Experimental~\cite{Moddeman1971} and calculated normal Auger spectra of core-ionized  \ce{CO2} at the carbon (left panels) and oxygen (right panels) K-edges. The theoretical spectra were obtained by applying Gaussian broadening with a full width at half maximum (FWHM) of 0.7~eV to the stick spectra. Gray and blue sticks represent singlet and triplet decay channels, respectively. The OCA spectra are taken from Ref.~\citenum{Skomorowski2025}, while the present results were obtained using the LS-CW-IC model (for C edge) and the LS-PW-IC model (for O edge).
 }
\label{fig:KLL-co2-c}
\end{figure*}

Finally, we compare the calculated Auger spectra with experiment. Figure~\ref{fig:KLL-co2-c} presents the experimental spectra alongside those obtained using the LS and OCA approaches for both the C and O K-edges. Despite the noticeable differences in absolute partial widths discussed above, the LS and OCA spectra exhibit remarkably similar overall shapes. This agreement arises because the relative widths, particularly those of the dominant decay channels, are very similar in the two approaches. For the oxygen K-edge, the calculated spectra are nearly indistinguishable, whereas for the carbon K-edge, more visible differences occur in the binding-energy range of 62--75~eV. Nevertheless, the two approaches yield almost identical spectral profiles in the region containing the dominant features. Comparison with experiment indicates that the remaining discrepancies cannot be resolved simply by improving the continuum orbital representation. Instead, further improvements likely require a more accurate treatment of vibrational broadening and possibly higher-order electron correlation beyond EOM-CCSD.

The results for \ce{CO2} allow us to draw several general conclusions regarding the application of the present LS approach with an SGTO continuum basis to molecular KLL Auger decay. First, the influence of the scattering potential on the partial widths is less pronounced in \ce{CO2} than in Ne or \ce{H2O}. This may suggest that the sensitivity of Auger decay widths to the continuum representation decreases with increasing molecular size, potentially helping to explain the relative success of the simple PW approximation in reproducing experimental Auger spectra of medium-sized molecules such as benzene and oxazole~\cite{Jayadev:23, oxazole:2022}. Therefore, the LS-PW formulation should be sufficient to capture the dominant effects of the scattering potential, avoiding the additional computational cost associated with the LS-CW formulation and the inclusion of interchannel coupling, which becomes prohibitively expensive for larger systems.

Second, the continuum basis-set analysis indicates that an accurate description can be achieved using a relatively compact basis localized on the core-ionized atom, with the angular-momentum expansion restricted to the same partial waves as required in the isolated atom.
The weak sensitivity to continuum functions on neighboring atoms may also explain the good performance of the OCA method, which effectively restricts the continuum representation to the core-ionized atomic center. Finally, the comparison with experiment suggests that further improvements in the predicted Auger spectra will require a more accurate treatment of nuclear dynamics, satellite decay channels, and higher-order electron correlation, rather than additional refinement of the continuum representation alone. 

\subsection{L$_1$MM Auger decay of Ar}

As a final example, we consider the decay of an L$_1$ (2$s^{-1}$) core hole in the Ar atom, another well-established benchmark system for Auger decay that has been extensively studied both experimentally and theoretically\cite{Mehlhorn1968, Werme1973, Chen1981, Pulkkinen1996, Samson1997, Liu2021}. Core-hole decay in Ar is considerably more complex than in second-row atoms such as C, N, O, and Ne, owing to the possibility of multistep decay cascades and multiple-electron emission. In particular, the dominant decay pathways of an L$_1$ core hole involve an initial Coster--Kronig transition followed by further autoionization\cite{Mehlhorn1968, Liu2021}, ultimately producing \ce{Ar^{3+}} ions. Here, however, we focus exclusively on normal L$_1$MM Auger decay, which produces stable \ce{Ar^{2+}} final states that do not undergo further autoionization. The availability of high-quality experimental and theoretical data\cite{Lablanquie2011, Liu2021} for these channels provides an opportunity to further assess the accuracy of the present SGTO Lippmann--Schwinger approach.

\begin{table*}[t!]
\caption{
Auger electron energies (eV) and partial widths $\Gamma_i$ (meV) for L$_1$MM transitions in core-ionized Ar, calculated using the cc-pVQZ / $[29(spd);1.6,13000]$ basis sets with different continuum orbital models, compared with selected theoretical results from the literature.}
\label{tab:Ar_L1MM}
\centering
\scriptsize
\setlength{\tabcolsep}{3.2pt}
\renewcommand{\arraystretch}{1.05}
\resizebox{\textwidth}{!}{%
\begin{tabular}{lccccccc|ccc}
\hline
\multirow{2}{*}{Channel}
& \multirow{2}{*}{Auger energy}
& \multicolumn{6}{c|}{Present continuum model}
& \multicolumn{3}{c}{Other theory} \\
\cline{3-8}
\cline{9-11}
&
& PW
& Proj-PW
& LS-PW
& LS-CW
& LS-PW-IC
& \multicolumn{1}{c|}{LS-CW-IC}
& MCDF\textsuperscript{\emph{a}}
& CBF\textsuperscript{\emph{b}}
& CS\textsuperscript{\emph{c}}
\\

\hline
$^1$D$(3p^{-2})$ & 280.2 & 6.7 & 6.7 & 0.2 & 0.3 & 1.0 & 1.0 & -- & -6.5 & -6.4 \\
$^1$S$(3p^{-2})$ & 277.8 & 0.6 & 0.6 & 0.5 & 0.6 & 0.5 & 0.5 & 0.7\textsuperscript{\emph{d}} & 4.0 & 4.1 \\
$^1$P$(3s^{-1}3p^{-1})$ & 261.7 & 2.8 & 2.8 & 27.9 & 28.9 & 27.7 & 28.7 & 15.3 & 45.1 & 46.6 \\
$^1$S$(3s^{-2})$ & 247.6 & 19.3 & 19.3 & 13.6 & 12.8 & 13.1 & 12.3 & 8.2 & 13.6 & 13.7 \\
$^3$P$(3p^{-2})$ & 281.9 & 0.0 & 0.0 & 0.0 & 0.0 & 0.0 & 0.0 & -- & 0.0 & 0.0 \\
$^3$P$(3s^{-1}3p^{-1})$ & 266.9 & 2.3 & 2.3 & 36.7 & 37.2 & 36.0 & 36.5 & 30.6\textsuperscript{\emph{e}} & 19.1 & 18.9 \\
\hline
$\sum \Gamma_i$ &  & 31.8 & 31.8 & 78.9 & 79.8 & 78.2 & 79.1 & 55.0 & 75.3 & 77.0 \\
\hline
\end{tabular}%
}\\
\textsuperscript{\emph{a}} Multiconfiguration Dirac-Fock with numerical continuum orbital~\cite{Liu2021} \\
\textsuperscript{\emph{b}} Complex basis function EOM-IP-CCSD  with the aug-cc-pCV5Z basis set augmented by 8  spd shells\cite{Drennhaus2024}\\
\textsuperscript{\emph{c}} Complex scaling EOM-IP-CCSD  with the aug-cc-pCV5Z basis set\cite{Drennhaus2024}\\
\textsuperscript{\emph{d}} Combined value of $3p^{-2}$ ($^1$S, $^1$D, $^3$P)\\
\textsuperscript{\emph{e}} Combined value of $^3$P for $J=0,1,2$ respectively
\end{table*}
Table~\ref{tab:Ar_L1MM} compares the partial widths calculated using different continuum representations for the main two-hole ($2h$) channels of Ar L$_1$MM Auger decay, together with selected theoretical results from the literature. As in Ne, incorporating the scattering potential through the Lippmann--Schwinger equation has a large effect on the partial widths, with PW and LS-PW results differing by more than an order of magnitude for several channels. In contrast, the effects of the LS-CW formulation and interchannel coupling are very small.
The distribution of partial widths also differs markedly from that of Ne KLL decay. In Ar, the dominant contributions arise from the $3s^{-1}3p^{-1}$ configurations, with comparable widths for the corresponding singlet and triplet channels. This contrasts with Ne, where the $2p^{-2}$ singlet channel dominates and triplet contributions are much smaller.

\begin{table*}[t!]
\caption{
Effect of triple excitations on the double ionization potentials (EOM-DIP, eV), squared norms of the two-hole EOM-DIP amplitudes ($\|R_2\|^2$), and partial Auger decay widths ($\Gamma_i$, meV) for the L$_1$MM decay channels of core-ionized Ar. The present CCSD widths were calculated using the LS-CW-IC model with cc-pVQZ / $[29(spd);1.6,13000]$ basis sets, while the CCSDT$^*$ widths were estimated using Eq.~\eqref{eq:ccsdt}.}
\label{tab:Ar_L1MM_2}
\centering
\scriptsize
\begin{tabular}{lccccccc}
\hline
\multirow{2}{*}{Channel}& \multicolumn{2}{c}{EOM-DIP} & \multicolumn{2}{c}{$||R_2||^2$} & \multicolumn{3}{c}{$\Gamma_i$}  \\
\cmidrule(r){2-3}\cmidrule(l){4-5}\cmidrule(l){6-8}
 & CCSD & CCSDT & CCSD & CCSDT & CCSD & CCSDT$^*$ & MCDF\cite{Liu2021}\\
\hline
$^1$D$(3p^{-2})$         & 45.3 & 44.9 & 0.9082 & 0.8676 & 1.0  & 1.0  & -- \\
$^1$S$(3p^{-2})$         & 47.8 & 47.4 & 0.9288 & 0.8935 & 0.5  & 0.5  & 0.7 \\
$^1$P$(3s^{-1}3p^{-1})$  & 63.9 & 61.7 & 0.7116 & 0.4668 & 28.7 & 18.8 & 15.3 \\
$^1$S$(3s^{-2})$         & 78.0 & 75.0 & 0.6993 & 0.4770 & 12.3 & 8.4  & 8.2 \\
$^3$P$(3p^{-2})$         & 43.7 & 43.2 & --     & --     & 0.0  & 0.0  & -- \\
$^3$P$(3s^{-1}3p^{-1})$  & 58.7 & 57.4 & 0.8097 & 0.6870 & 36.5 & 31.0 & 30.6 \\
\hline
\multicolumn{8}{c}{Satellites\textsuperscript{\emph{a}}}\\
\hline
$^1$P$(3p^{-3}3d$) & 75.6 & 71.0 & 0.1676 & 0.3009 & 5.8  & 10.4 & -- \\
$^1$P$(3p^{-3}4s$) & 76.6 & 69.5 & 0.0578 & 0.0417 & 2.2  & 1.6  & -- \\
$^1$P$(3p^{-3}3d$) & 79.7 & 73.7 & 0.0037 & 0.0165 & 0.2  & 0.8  & -- \\
$^1$P$(3p^{-3}3d$) & 86.5 & 79.4 & 0.0204 & 0.0447 & 0.7  & 1.5  & -- \\
$^3$P$(3p^{-3}4s$) & 75.3 & 69.1 & 0.0047 & 0.0156 & 0.2  & 0.6  & -- \\
$^3$P$(3p^{-3}3d$) & 72.9 & 67.8 & 0.0146 & 0.0041 & 0.6  & 0.2  & -- \\
$^3$P$(3p^{-3}3d$) & 76.2 & 70.4 & 0.0916 & 0.1269 & 3.8  & 5.2  & -- \\
\hline
\end{tabular}\\
\textsuperscript{\emph{a}} Calculated using cc-pVDZ + 2d bound-state basis set and neglecting interchannel coupling. \\
\end{table*}

Table~\ref{tab:Ar_L1MM} also includes results obtained using complex basis functions (CBF) and complex scaling (CS) within the EOM-IP-CCSD framework, as well as multiconfiguration Dirac--Hartree--Fock (MCDHF) calculations employing numerical orbitals, which can serve as a reference. The present results show reasonable agreement with MCDHF, although the latter predicts a substantially smaller width for the $^1$P$(3s^{-1}3p^{-1})$ channel. The CBF and CS results exhibit larger discrepancies, including unphysical negative partial widths for the $^1$D$(3p^{-2})$ channel, reflecting limitations of the procedure used to decompose the total decay width into individual channel contributions.

Since neither interchannel coupling nor the choice of LS-PW versus LS-CW substantially changes the present results, the remaining discrepancies with MCDHF are likely associated with the bound-state electronic-structure description. Indeed, Ref.~\citenum{Liu2021} demonstrated significant mixing of the nominal $2h$ states with three-hole--one-particle ($3h1p$) satellite configurations involving excitations into the $3d$ and $4s$ orbitals. Such configurations cannot be adequately described within EOM-DIP-CCSD. To assess the importance of higher-order correlation effects, we therefore performed EOM-DIP-CCSDT calculations, which include triple excitations in the coupled-cluster operator and $3h1p$ configurations in the EOM operator.

A fully consistent EOM-CCSDT treatment of Auger decay would also require corresponding modifications in the initial core-ionized state and Dyson transition amplitudes, which are beyond the scope of this work. Instead, we estimate the effect of higher-order correlation on the partial widths by exploiting the predominantly one-hole character of the initial $\mathrm{Ar}^{+}(2s^{-1})$ state. Under this approximation, the Auger partial widths are assumed to scale with the squared norm of the $2h$ component of the final-state wave function:
\begin{equation}
\Gamma^{\mathrm{CCSDT}}\approx
\Gamma^{\mathrm{CCSD}}
\frac{\left\|R_{2}^{\mathrm{DIP\text{-}CCSDT}}\right\|^2}
{\left\|R_{2}^{\mathrm{DIP\text{-}CCSD}}\right\|^2}.
\label{eq:ccsdt}
\end{equation}
\begin{figure*}[th!]
      \centering
      \includegraphics[width=0.93\textwidth]{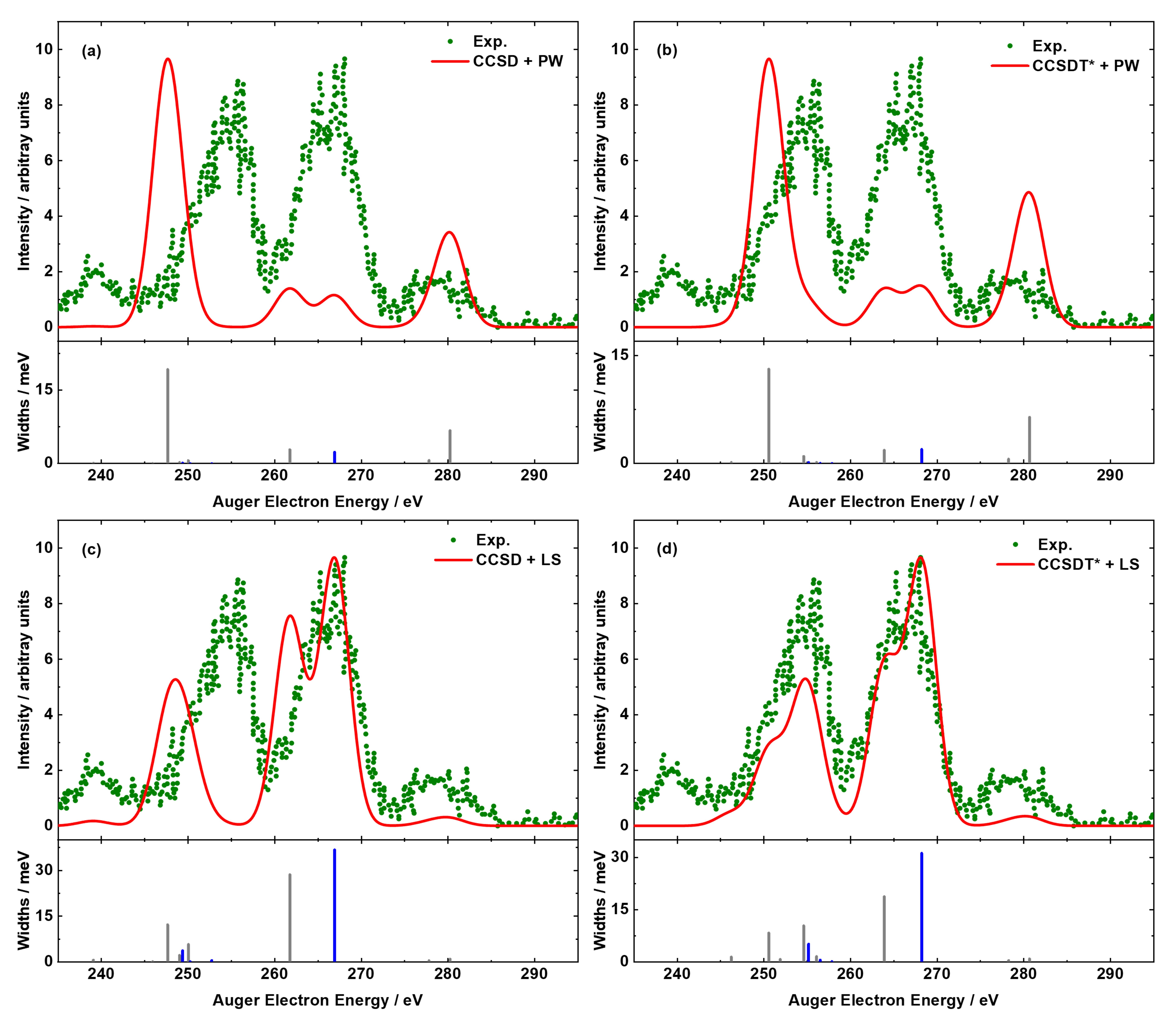}
      \caption{Comparison of calculated and experimental\cite{Lablanquie2011} Ar L$_1$MM Auger spectra. The upper panels (a,b) show EOM-CCSD and CCSDT*-corrected results obtained using the plane-wave (PW) continuum model, while the lower panels (c,d) show the corresponding results using the LS-CW-IC model. Gray and blue sticks represent singlet and triplet decay channels, respectively. The theoretical spectra were generated by applying a uniform Gaussian broadening of 1.7~eV (FWHM) to the stick spectra to approximately match the experimental resolution.}
      \label{fig:Ar_L1MM}
  \end{figure*}
The resulting estimates, which approximately account for the effect of triple excitations on the final dicationic states, are presented in Table~\ref{tab:Ar_L1MM_2}. In addition to the main two-hole (2h) channels, Table~\ref{tab:Ar_L1MM_2} includes also satellite states lying between the dominant $3s^{-1}3p^{-1}$ and $3s^{-2}$ configurations. Inspection of the results reveals a substantial effect of configuration mixing, which reduces the squared norm of the $2h$ component $\|R_2\|^2$ by more than 30\% for the $^1$P$(3s^{-1}3p^{-1})$ channel. Remarkably, the CCSDT-corrected partial widths are now in excellent agreement with the reference MCDHF values for all $2h$-dominated channels. This agreement suggests that the effects of higher-order electron correlation can be approximately incorporated at a substantially lower computational cost than a fully consistent EOM-CCSDT treatment, without significantly compromising the accuracy of the resulting widths.

Finally, Fig.~\ref{fig:Ar_L1MM} compares the calculated Auger spectra with the experiment~\cite{Lablanquie2011}. The different panels show spectra obtained using either PW or LS-CW-IC continuum orbitals, combined with EOM-CCSD or EOM-CCSDT descriptions of the bound states. The experimental spectrum is dominated by two broad features: the first, around 265--270~eV, arises mainly from the $3s^{-1}3p^{-1}$ channels, while the second, around 250--260~eV, originates from the $3s^{-2}$ channel and satellite states dominated by $3p^{3}3d$ configurations. The upper panels demonstrate that the PW approximation fails to reproduce the experimental branching ratios. The lower panels show that the LS-CW-IC continuum orbitals substantially improve the relative intensities, while the inclusion of satellite states and higher-order correlation effects through EOM-CCSDT is essential for achieving satisfactory agreement with experiment in both peak positions and relative intensities. Thus, a quantitatively reliable description of the Ar L$_1$MM Auger spectrum requires not only an accurate representation of the scattering continuum but also a correlated treatment of the final dicationic states beyond the lowest two-hole configurations.

\section{Conclusions and outlook}
The present work addresses a longstanding challenge in Gaussian-based quantum chemistry by introducing a practical approach for describing continuum electron within the same computational framework used for bound electronic states. By constructing scattering continuum orbital directly in a Gaussian-type basis, the methodology enables their seamless integration with established electronic-structure methods while retaining the essential physical requirements of continuum states.

The proposed methodology is based on solving the Lippmann--Schwinger (LS) equation in a finite auxiliary basis of spherical Gaussian-type orbitals (SGTOs). It provides scattering continuum orbital that account for the interaction of the outgoing electron with the atomic or molecular potential while incorporating the appropriate outgoing-wave boundary conditions. An important feature of the approach is its systematic improvability. Starting from a plane-wave description, the continuum representation can be refined by including the scattering potential through either a plane-wave (LS-PW) or Coulomb-wave (LS-CW) reference state, followed by the inclusion of interchannel coupling. This hierarchy allows the level of approximation to be adjusted to the physical problem and the required accuracy. We have also examined practical strategies for constructing efficient auxiliary SGTO basis sets. In particular, plane-wave calculations provide a useful starting point for identifying the Gaussian functions most relevant to a given process, while the analysis of free-particle Green's-function matrix elements offers additional guidance for selecting suitable exponents at a given continuum-electron energy.

As the first application of the methodology, we investigated normal Auger decay in atoms and small molecules. The localized character of the Auger transition makes it a particularly suitable test of the continuum representation. Calculations for Ne and Ar demonstrate substantial improvements over the plane-wave approximation, with partial decay widths in close agreement with available experimental data and high-level theoretical references. The molecular benchmarks, \ce{H2O} and \ce{CO2},  demonstrate that the approach can provide accurate absolute Auger widths while retaining the flexibility of a standard Gaussian-basis electronic-structure framework. At the same time, comparison with experimental spectra reveals limitations arising from the description of nuclear dynamics, satellite transitions, and higher-order electron correlation, which cannot be overcome by improving the continuum orbitals alone.

Our molecular benchmarks show the LS-PW formulation captures the dominant effects of the scattering potential on Auger widths, while the additional corrections from LS-CW and interchannel coupling are generally smaller. Thus, LS-PW should provide a sufficiently accurate continuum description for many Auger decay applications, avoiding the prohibitive cost of interchannel coupling in larger systems. The molecular benchmarks also demonstrate that compact SGTO basis sets localized on the core-ionized atom are sufficient for an accurate continuum representation. This finding is consistent with the good performance of the one-center approximation (OCA), which reproduces the relative widths of the dominant Auger channels despite noticeable differences in their absolute values.

While the present benchmarks employ EOM-CCSD-based bound-state wave functions, the continuum methodology is general. Its formulation in terms of electronic state and transition-state densities allows integration with other methods, including higher-level coupled-cluster and DFT/TDDFT approaches. The Ar L$_1$MM results illustrate the importance of this flexibility: accurate continuum orbitals alone cannot reproduce all major spectral features when satellite states and strong configuration mixing require a more complete treatment of electron correlation. A key advantage of the present framework is thus the ability to improve the continuum and many-electron descriptions independently and systematically assess their contributions to the calculated observables.

Overall, this work establishes a practical, systematically improvable framework for treating continuum electron within Gaussian-based quantum chemistry, combining scattering boundary conditions with correlated electronic-structure methods. Although demonstrated here for normal Auger decay, the formulation is broadly applicable. Extensions to resonant Auger decay, Coster--Kronig transitions, photoionization, and electron-impact processes will be presented in forthcoming publications. These studies will also examine whether the conclusions regarding the SGTO continuum basis and the different LS models remain valid for other continuum-electron processes.

\begin{acknowledgements}
We acknowledge financial support from the Polish National Agency for Academic Exchange through Polish Returns Programme (W.S.) and Ulam fellowship (D.M.), and from the National Science Centre, Poland (SONATA 2020/39/D/ST4/03048). We gratefully acknowledge Polish high-performance computing infrastructure PLGrid (HPC Center: ACK Cyfronet AGH) for providing computer facilities and support within computational grant no. PLG/2024/017494. 
\end{acknowledgements}



\bibliography{achemso-demo}
\end{document}